\documentclass[aps,onecolumn,10pt]{revtex4}
\usepackage{amsmath}
\usepackage{amssymb}
\usepackage{hyperref}
\usepackage{graphicx}
\usepackage{float}
\usepackage{slashed}
\numberwithin{equation}{section}

\hypersetup{colorlinks=true} 
\begin{document}
\allowdisplaybreaks
\setcounter{equation}{0}
 
\title{Equivalence of light-front quantization and instant-time quantization}

\author{Philip D. Mannheim}
\affiliation{Department of Physics, University of Connecticut, Storrs, CT 06269, USA \\
philip.mannheim@uconn.edu}

\date{June 15 2020}

\begin{abstract}

Commutation or anticommutation relations quantized at equal instant time and commutation or anticommutation relations quantized at equal light-front time not  only cannot be transformed into each other, they take completely different forms. While they would thus appear to describe different theories, we show that this is not in fact the case. By looking not at equal times but at unequal times, we show that unequal instant-time commutation or anticommutation relations are completely equivalent to unequal light-front time commutation or anticommutation relations. Light-front quantization and instant-time quantization are thus the same and thus describe the same theory, with it being only the restriction to equal times that makes them look different. However for fermions there is a caveat, as the light-front anticommutation relations involve projection operators acting on the fermion fields. Nonetheless, not only can one  still derive fermion unequal light-front time anticommutators starting from unequal instant-time ones, one can even derive unequal instant-time fermion anticommutators starting from unequal light-front time anticommutators even though the fermion projection operators that are relevant in the light-front case are not invertible.  To establish the equivalence for gauge fields we present a quantization procedure that does not involve the zero-mode singularities that are commonly encountered in light-front gauge field studies. We also study time-ordered products of fields, and again show the equivalence despite the fact that there are additional terms in the fermion light-front case.  We establish our results first  for free theories, and then to all orders in interacting theories though comparison of the instant-time and light-front Lehmann representations. Finally, we compare instant-time Hamiltonians and light-front Hamiltonians, and show that in the instant-time rest frame they give identical results.

\end{abstract}

\maketitle

\section{Introduction}
\label{S1}

In quantum field theory various choices of quantization are considered. The most common choice is to take commutation relations of pairs of fields at equal instant time $x^0$ to be specific singular c-number functions. Thus for a free scalar field with action 
\begin{eqnarray}
I_S&=&\int dx^0dx^1dx^2dx^3\tfrac{1}{2}\left[(\partial_0\phi)^2-(\partial_1\phi)^2-(\partial_2\phi)^2-(\partial_3\phi)^2-m^2\phi^2\right]
\label{1.1a}
\end{eqnarray}
for instance,  one identifies a canonical conjugate $\delta I_S/\delta \partial_0\phi=\partial^0\phi=\partial_0\phi$ (one can of course add on interaction terms to $I_S$, but as long as they contain no derivatives they do not affect the identification of the canonical conjugate), and then quantizes the theory according to the equal instant-time canonical commutation relation
\begin{eqnarray}
[\phi(x^0,x^1,x^2,x^3), \partial_0\phi(x^0,y^1,y^2,y^3)]=i\delta(x^1-y^1)\delta(x^2-y^2)\delta(x^3-y^3).
\label{1.2a}
\end{eqnarray}
In light-front quantization (see e.g. \cite{Brodsky:1997de} for a review) one introduces coordinates $x^{\pm}=x^0\pm x^3$, a line element $g_{\mu\nu}x^{\mu}x^{\nu}=x^+x^--(x^1)^2-(x^2)^2$  with $(-g)^{1/2}=1/2$, and a free scalar field action of the form
\begin{eqnarray}
I_S&=&\tfrac{1}{2}\int dx^+dx^1dx^2dx^-\tfrac{1}{2}\left[2\partial_+\phi\partial_-\phi+2\partial_-\phi\partial_+\phi-(\partial_1\phi)^2-(\partial_2\phi)^2-m^2\phi^2\right].
\label{1.3a}
\end{eqnarray}
One identifies a canonical conjugate $(-g)^{-1/2}\delta I_S/\delta \partial_+\phi=\partial^+\phi=2\partial_-\phi$, and quantizes the theory according to the equal light-front time $x^+$ commutation relation (see e.g. \cite{Neville1971} and more recently \cite{Mannheim2019a}) 
\begin{eqnarray}
[\phi(x^+,x^1,x^2,x^-), 2\partial_-\phi(x^+,y^1,y^2,y^-)]=i\delta(x^1-y^1)\delta(x^2-y^2)\delta(x^--y^-).
\label{1.4a}
\end{eqnarray}
As written, (\ref{1.4a}) is already conceptually different from (\ref{1.2a}) since the light-front conjugate is $2\partial_-\phi$ and not $2\partial_+\phi$, i.e., not the derivative with respect to the light-front time, while the instant-time conjugate $\partial_0\phi$ is the derivative with respect to the instant time. Since $\phi(x^+,x^1,x^2,x^-)$ and $\partial_-\phi(x^+,y^1,y^2,y^-)$ are not at the same $x^-$, (\ref{1.4a}) can be integrated to
\begin{eqnarray}
[\phi(x^+,x^1,x^2,x^-), \phi(x^+,y^1,y^2,y^-)]=-\frac{i}{4}\epsilon(x^--y^-)\delta(x^1-y^1)\delta(x^2-y^2),
\label{1.5a}
\end{eqnarray}
where $\epsilon(x)=\theta(x)-\theta(-x)$. Since the analog instant-time commutation relation  is given by
\begin{eqnarray}
[\phi(x^0,x^1,x^2,x^3), \phi(x^0,y^1,y^2,y^3)]=0,
\label{1.6a}
\end{eqnarray}
instant-time and light-front time quantization appear to be quite different. Similar concerns affect gauge field commutators.

For fermions instant-time and light-front time quantization again appear to be quite different, and in fact even more so. In instant-time quantization the free fermionic Dirac action is of the form
\begin{eqnarray}
I_D=\int dx^0dx^1dx^2dx^3\bar{\psi}[i(\gamma^{0}\partial_{0}+\gamma^{1}\partial_{1}+\gamma^{2}\partial_{2}+\gamma^{3}\partial_{3})-m]\psi. 
\label{1.7a}
\end{eqnarray}
The canonical conjugate of $\psi$ is $i\psi^{\dagger}$, and the canonical anticommutation relations are of the form
\begin{align}
&\Big{\{}\psi_{\alpha}(x^0,x^1,x^2,x^3),\psi_{\beta}^{\dagger}(x^0,y^1,y^2,y^3)\Big{\}}
=\delta_{\alpha\beta}\delta(x^1-y^1)\delta(x^2-y^2)\delta(x^3-y^3),
\nonumber\\
&\Big{\{}\psi_{\alpha}(x^0,x^1,x^2,x^3),\psi_{\beta}(x^0,y^1,y^2,y^3)\Big{\}}=0.
\label{1.8a}
\end{align}
For the light-front case we set 
\begin{eqnarray}
\partial_0=\frac{\partial}{\partial x^+}+\frac{\partial}{\partial x^-}=\partial_++\partial_-,\quad 
\partial_3=\frac{\partial}{\partial x^+}-\frac{\partial}{\partial x^-}=\partial_+-\partial_-,
\label{1.9a}
\end{eqnarray}
and obtain
\begin{eqnarray}
\gamma^0\partial_0+\gamma^3\partial_3=(\gamma^0+\gamma^3)\partial_++(\gamma^0-\gamma^3)\partial_-
=\gamma^+\partial_++\gamma^-\partial_-,
\label{1.10a}
\end{eqnarray}
with (\ref{1.10a}) serving to define $\gamma^{\pm}=\gamma^0\pm \gamma^3$. In terms of $\gamma^+$ and $\gamma^-$ the Dirac action takes the form
\begin{eqnarray}
I_D=\tfrac{1}{2}\int dx^+dx^1dx^2dx^-\psi^{\dagger}\gamma^0[i(\gamma^+\partial_++\gamma^-\partial_-+\gamma^1\partial_1+\gamma^2\partial_2)-m]\psi.
\label{1.11a}
\end{eqnarray}
With this action the light-front time canonical conjugate of $\psi$ is $i\psi^{\dagger}\gamma^0\gamma^+$. 
In the construction of the light-front fermion sector we find a rather sharp distinction with the instant-time fermion sector. First, unlike $\gamma^0$ and $\gamma^3$, which obey $(\gamma^0)^2=1$, $(\gamma^3)^2=-1$, $\gamma^+$ and $\gamma^-$ obey $(\gamma^+)^2=0$, $(\gamma^-)^2=0$, to thus both be non-invertible divisors of zero. Secondly, the quantities 
\begin{eqnarray}
\Lambda^{+}=\tfrac{1}{2}\gamma^0\gamma^{+}=\tfrac{1}{2}(1+\gamma^0\gamma^3),\quad \Lambda^{-}=\tfrac{1}{2}\gamma^0\gamma^{-}=\tfrac{1}{2}(1-\gamma^0\gamma^3)
\label{1.12a}
\end{eqnarray}
obey 
\begin{eqnarray}
\Lambda^{+}+\Lambda^{-}=I,\quad(\Lambda^{+})^2=\Lambda^{+}=[\Lambda^+]^{\dagger},\quad (\Lambda^{-})^2=\Lambda^{-}=[\Lambda^{-}]^{\dagger},\quad \Lambda^{+}\Lambda^{-}=0.
\label{1.13a}
\end{eqnarray}
We recognize (\ref{1.13a}) as a projector algebra, with $\Lambda^{+}$ and $\Lambda^{-}$ thus being non-invertible projection operators. Given the projector algebra we identify $\psi_{(+)}=\Lambda^+\psi$, $\psi_{(-)}=\Lambda^-\psi$ (respectively known as good and bad fermions in the light-front literature), and thus identify the conjugate of $\psi$ as $2i\psi_{(+)}^{\dagger}$, where $\psi^{\dagger}_{(+)}$ denotes $[\psi^{\dagger}]_{(+)}=\psi^{\dagger}\Lambda^+$, which is equal to $[\Lambda^+\psi]^{\dagger}=[\psi_{(+)}]^{\dagger}$ since $\Lambda^+$ is Hermitian. Since the conjugate is a good fermion, in the anticommutator of $\psi$ with its conjugate only the good component of $\psi$ will contribute since $\Lambda^+\Lambda^-=0$, with the equal light-front time canonical anticommutator being found to be of the form (\cite{Chang1973} and more recently \cite{Mannheim2019a})
\begin{eqnarray}
&&\big{\{}[\psi_{(+)}]_{\alpha}(x^+,x^1,x^2,x^-),[\psi_{(+)}^{\dagger}]_{\beta}(x^+,y^1,y^2,y^-)\big{\}}=\Lambda^+_{\alpha\beta}\delta(x^--y^-)\delta(x^1-y^1)\delta(x^2-y^2).
\label{1.14a}
\end{eqnarray}

In this construction the bad fermion $\psi_{(-)}$ has no canonical conjugate and is thus not a dynamical variable. To understand this in more detail we manipulate the Dirac equation $(i\gamma^+\partial_++i\gamma^-\partial_-+i\gamma^1\partial_1+i\gamma^2\partial_2-m)\psi=0$. We first multiply on the left  by $\gamma^0$ to obtain
\begin{eqnarray}
2i\partial_+\psi_{(+)}+2i\partial_-\psi_{(-)}+i\gamma^0(\gamma^1\partial_1+\gamma^2\partial_2)\psi-m\gamma^0\psi=0.
\label{1.15a}
\end{eqnarray}
Next we multiply (\ref{1.15a}) by $\Lambda^-$ and also multiply it by $\Lambda^+$ to obtain
\begin{eqnarray}
2i\partial_-\psi_{(-)}&=&[-i\gamma^0(\gamma^1\partial_1+\gamma^2\partial_2)+m\gamma^0]\psi_{(+)},
\quad
2i\partial_+\psi_{(+)}=[-i\gamma^0(\gamma^1\partial_1+\gamma^2\partial_2)+m\gamma^0]\psi_{(-)}.
\label{1.16a}
\end{eqnarray}
Since the $\partial_-\psi_{(-)}$ equation contains no time derivatives, $\psi_{(-)}$ is thus a constrained variable, consistent with it having no conjugate. Through the use of the inverse propagator $(\partial_-)^{-1}(x^-)=\epsilon(x^-)/2$ we can rewrite the $\partial_-\psi_{(-)}$ equation in (\ref{1.15a}) as  
\begin{align}
\psi_{(-)}(x^+,x^1,x^2,x^-)&=\frac{1}{4i}\int du^-\epsilon(x^--u^-)[-i\gamma^0(\gamma^1\partial_1+\gamma^2\partial_2)+m\gamma^0]\psi_{(+)}(x^+,x^1,x^2,u^-),
\nonumber\\
[\psi_{(-)}]^{\dagger}&=\frac{i}{4}\int du^-\epsilon(x^--u^-)[i\partial_1[\psi_{(+)}]^{\dagger}\gamma^0\gamma^1+
i\partial_2[\psi_{(+)}]^{\dagger}\gamma^0\gamma^2+m[\psi_{(+)}]^{\dagger}\gamma^0],
\label{1.17a}
\end{align}
and recognize $\psi_{(-)}$ as obeying a constraint condition that is  nonlocal. It is because $\psi_{(-)}$ obeys such a nonlocal constraint  that it is known as a bad fermion. Since it is a constrained variable it does not appear in any fundamental anticommutation relation. Nonetheless, one can still use (\ref{1.14a}) and (\ref{1.17a}) to construct a bad fermion bad fermion $\big{\{}\psi_{(-)},\psi^{\dagger}_{(-)}\big{\}}$ anticommutator. In this way we obtain \cite{Mannheim2019a,Mannheim2020a}
\begin{align}
&\Big{\{}\frac{\partial}{\partial x^-}\psi_{\alpha}^{(-)}(x^+,x^1,x^2,x^-),\frac{\partial}{\partial y^-}[\psi_{(-)}^{\dagger}]_{\beta}(x^+,y^1,y^2,y^-)\Big{\}}
\nonumber\\
&=\frac{1}{4}\Lambda^-_{\alpha\beta}\left[-\frac{\partial}{\partial x^1}\frac{\partial}{\partial x^1}-\frac{\partial}{\partial x^2}\frac{\partial}{\partial x^2}+m^2\right]\delta(x^--y^-)\delta(x^1-y^1)\delta(x^2-y^2),
\label{1.18a}
\end{align}
which integrates to 
\begin{align}
&\Big{\{}\psi_{\alpha}^{(-)}(x^+,x^1,x^2,x^-),[\psi_{(-)}^{\dagger}]_{\beta}(x^+,y^1,y^2,y^-)\Big{\}}
\nonumber\\
&=\frac{1}{16}\Lambda^-_{\alpha\beta}\left[-\frac{\partial}{\partial x^1}\frac{\partial}{\partial x^1}-\frac{\partial}{\partial x^2}\frac{\partial}{\partial x^2}+m^2\right]
\int du^-\epsilon(x^--u^-)\epsilon(y^--u^-)\delta(x^1-y^1)\delta(x^2-y^2).
\label{1.19a}
\end{align}
As we see, the equal $x^+$ bad fermion sector $\Big{\{}\psi_{\alpha}^{(-)}(x^+,x^1,x^2,x^-),[\psi_{(-)}^{\dagger}]_{\beta}(x^+,y^1,y^2,y^-)\Big{\}}$ anticommutator is non-vanishing, with its nonlocal nature being apparent. However this non-locality is restricted to the light cone since with $x^+=y^+$, $x^1=y^1$, $x^2=y^2$ the quantity  $(x^+-y^+)(x^--y^-)-(x^1-y^1)^2-(x^2-y^2)^2$ is zero for any value of $x^--y^-$.

By this same procedure we can construct the good fermion bad fermion $\Big{\{}\psi_{\mu}^{(+)}(x^+,x^1,x^2,x^-),[\psi_{(-)}^{\dagger}]_{\nu}(x^+,y^1,y^2,y^-)\Big{\}}$ anticommutator and its Hermitian conjugate as well, and obtain \cite{Mannheim2019a,Mannheim2020a}
\begin{align}
&\Big{\{}[\psi_{(+)}]_{\nu}(x),[\psi^{\dagger}_{(-)}]_{\sigma}(y)\Big{\}}
\nonumber\\
&
=\Big{\{}[\psi_{(+)}]_{\nu}(x^+,x^1,x^2,x^-),\tfrac{i}{4}\int du^-\epsilon(y^--u^-)[i\gamma^0(\gamma^1\partial_1^y+\gamma^2\partial_2^y)+m\gamma^0]_{\tau\sigma}[\psi^{\dagger}_{(+)}]_{\tau}(x^+,y^1,y^2,u^-)\Big{\}}
\nonumber\\
&=\tfrac{i}{4}\int du^-\epsilon(y^--u^-)\Lambda^+_{\nu\tau}[i\gamma^0(\gamma^1\partial_1^y+\gamma^2\partial_2^y)+m\gamma^0]_{\tau\sigma}
\delta(x^--u^-)\delta(x^1-y^1)\delta(x^2-y^2)\
\nonumber\\
&=\tfrac{i}{8}\epsilon(y^--x^-)[i(\gamma^-\gamma^1\partial_1^y+\gamma^-\gamma^2\partial_2^y)+m\gamma^-]_{\nu\sigma}\delta(x^1-y^1)\delta(x^2-y^2)
\nonumber\\
&=\tfrac{i}{8}\epsilon(x^--y^-)[i(\gamma^-\gamma^1\partial_1^x+\gamma^-\gamma^2\partial_2^x)-m\gamma^-]_{\nu\sigma}\delta(x^1-y^1)\delta(x^2-y^2),
\label{1.20a}
\end{align}
where $\partial^y_1$ denotes $\partial/\partial y^1$, etc. We note that unlike the bad fermion bad fermion anticommutator, the good fermion bad fermion anticommutator is still local.

As we see, the equal light-front time fermion sector anticommutators given in (\ref{1.14a}),  (\ref{1.18a}), (\ref{1.19a}) and (\ref{1.20a}) not only look different from their instant-time counterparts given in (\ref{1.8a}), because of the presence of the non-invertible good and bad projection operators they appear to be altogether inequivalent to their instant-time counterparts. Nonetheless, as shown in \cite{Mannheim2019a,Mannheim2019b} from a study of both Feynman diagrams and path integrals (a study that includes vacuum sector diagrams such as the one given in Fig. \ref{undressedtadpole} below), instant-time sector and light-front sector matrix elements of operators such as $\langle \Omega |T[\phi(x)\phi(y)]|\Omega\rangle$ or $\langle \Omega |T[\psi(x)\bar{\psi}(y)]|\Omega\rangle$ (as time ordered with $x^0$ or $x^+$) are in fact equal.  This therefore raises the question of how matrix elements of these very same operators could then actually be equal. And thus despite the seeming  differences both for commutators and anticommutators  it must be possible to establish that there nonetheless is an equivalence between the two quantization schemes at the operator level itself without needing to take matrix elements at all. This then is the objective of this paper. We shall meet this objective by looking not at equal-time commutators and anticommutators but at unequal-time ones, and show that it is only the restriction to equal times that makes the instant-time and light-front commutators and anticommutators look so different.  Central to our analysis will be the identification and treatment of singularities on the light cone that the unequal instant-time and unequal light-front time commutators and anticommutators possess.  Our results will be established first for free theories, and then to all orders in interacting theories though comparison of the instant-time and light-front all-order Lehmann representations. Finally, we compare instant-time Hamiltonians and light-front Hamiltonians, and show that in the instant-time rest frame they give identical results. 
\section{Unequal-Time Scalar Field Commutators}
\label{S2}

For the scalar field case  we note that in instant-time quantization one can use the equal-time commutation relation given in (\ref{1.2a}) and the wave equation $(\partial_{\mu}\partial^{\mu}+m^2)\phi=0$ associated with $I_S$ to make an on-shell Fock space expansion of $\phi(x)$ of the form
\begin{eqnarray}
\phi(x^0,\vec{x})= \int_{-\infty}^{\infty}dp_1\int_{-\infty}^{\infty}dp_2\int_{-\infty}^{\infty}dp_3\frac{1}{(2\pi)^{3/2}(2E_p)^{1/2}}[a(\vec{p})e^{-iE_p x^0+i\vec{p}\cdot\vec{x}}+a^{\dagger}(\vec{p})e^{+iE_p x^0-i\vec{p}\cdot\vec{x}}],
\label{2.1a}
\end{eqnarray}
where $E_p=(\vec{p}^2+m^2)^{1/2}$, and where the normalization of the creation and annihilation operator algebra, viz. $[a(\vec{p}),a^{\dagger}(\vec{q})]=\delta^3(\vec{p}-\vec{q})$, 
is fixed from the normalization of the canonical commutator given in (\ref{1.2a}). Given (\ref{2.1a}) one can evaluate the unequal instant-time commutation relation $i\Delta(IT;x)=[\phi(x^0,x^1,x^2,x^3), \phi(0,0,0,0)]$ between two free scalar fields, to obtain
\begin{eqnarray}
i\Delta(IT;x)&=&\int_{-\infty}^{\infty}dp_1\int_{-\infty}^{\infty}dp_2\int_{-\infty}^{\infty}dp_3\frac{1}{(2\pi)^3 2E_p}\left(e^{-i(E_px^0
+p_1x^1+p_2x^2+p_3x^3)}-e^{i(E_px^0+p_1x^1+p_2x^2+p_3x^3)}\right).
\label{2.2a}
\end{eqnarray}
We note that this unequal instant-time commutator is a c-number, and not a q-number. Since (\ref{2.1a}) itself is based on (\ref{1.2a}), (\ref{1.2a}) can be recovered from (\ref{2.2a}), as can (\ref{1.6a}). 

In the light-front case one can make an analogous Fock expansion for scalar fields, viz.
\begin{align}
\phi(x^+,x^1,x^2,x^-)&=\frac{2}{(2\pi)^{3/2}}\int_{-\infty}^{\infty}dp_1\int_{-\infty}^{\infty}dp_2\int_0^{\infty} \frac{dp_-}{(4p_-)^{1/2}}
\Big{[}e^{-i(F_p^2x^+/4p_-+p_-x^-+p_1x^1+p_2x^2)}a(\vec{p})
\nonumber\\
&+e^{i(F_p^2x^+/4p_-
+p_-x^-+p_1x^1+p_2x^2)}a^{\dagger}(\vec{p})\Big{]},
\label{2.3a}
\end{align}
where $F_p^2=(p_1)^2+(p_2)^2+m^2$, where the integration range for $p_-$ $(=p^+/2)$ is only over $p_-\geq 0$, and where the light-front $[a(\vec{p}),a^{\dagger}(\vec{p}^{~\prime})]$ commutator  is normalized to $[a(\vec{p}),a^{\dagger}(\vec{p}^{~\prime})]=(1/2)\delta(p_--p_-^{\prime})\delta(p_1-p_1^{\prime})\delta(p_2-p_2^{\prime})$, as fixed via the normalization of the equal light-front time canonical commutator given in (\ref{1.4a}). Given (\ref{2.3a}) we construct the unequal light-front time commutator $i\Delta(LF;x)=[\phi(x^+,x^1,x^2,x^-),\phi(0)]$, and obtain
\begin{align}
i\Delta(LF;x)&
=\frac{1}{4\pi^3}\int_{-\infty}^{\infty}dp_1\int_{-\infty}^{\infty}dp_2\int_{0}^{\infty}\frac{dp_-}{4p_-}
\left[e^{-i(F_p^2x^+/4p_-+p_-x^-+p_1x^1+p_2x^2)}-e^{i(F_p^2x^+/4p_-+p_-x^-+p_1x^1+p_2x^2)}\right].
\label{2.4a}
\end{align}
Since (\ref{2.4a}) itself is based on (\ref{1.4a}), both (\ref{1.4a}) and (\ref{1.5a}) can be recovered from (\ref{2.4a}). 

As noted in \cite{Harindranath1996}, if we could transform (\ref{2.2a}) into (\ref{2.4a}) we could then obtain equal light-front time commutation relations from unequal instant-time commutation relations. However, this cannot be done as is since in $i\Delta(IT;x)$ the variable $p_3$ ranges between $-\infty$ and $\infty$ while in $i\Delta(LF;x)$ the variable $p_-$ only ranges between $0$ and $\infty$. That $p_-$ could not be negative is due to the fact that in the on-shell light-front Fock expansion given in (\ref{2.3a}) there is a $1/(4p_-)^{1/2}$ term, and it has to be real. A second reason that (\ref{2.4a}) cannot be used as is is because it has a singularity at $p_-=0$ (viz. $p^+/2=0$), the zero-mode singularity that commonly appears in on-shell light-front studies and challenges them. As noted for instance in \cite{Mannheim2019a,Mannheim2019b} and as commented on further in Sec. \ref{S8} below, the way to handle on-shell zero-mode singularities such as these and give them a meaning is to go off shell. We shall thus rewrite $i\Delta(LF;x)$ as a contour integral in a complex $p_+$ ($=p^-/2$) plane, with no singularity problem then being found to occur. Doing this is necessary anyway since $i\Delta(IT;x)$ itself also has a singular structure, possessing $\epsilon(x^0)$ and $\delta(x^2)$ singularities, singularities that must then be reflected in and reproduced in $i\Delta(LF;x)$, with it precisely being the $x^2=0$ region where canonical commutators take support. As noted in \cite{Schwinger1949} evaluation of $i\Delta(IT;x)$ depends on whether $x^2$ is timelike, lightlike or spacelike, being given by \cite{Schwinger1949} (see also \cite{Schweber1964})
\begin{eqnarray}
i\Delta(IT;(x^2> 0)&=&\frac{i}{4\pi}\epsilon(x^0)\frac{mJ_1[m(x^2)^{1/2}]}{(x^2)^{1/2}},
\nonumber\\
i\Delta(IT;(x^2=0)&=&-\frac{i}{2\pi}\epsilon(x^0)\delta(x^2),
\nonumber\\
i\Delta(IT;(x^2<0)&=&0.
\label{2.5a}
\end{eqnarray}
Since commutators are evaluated on the light cone, central to our determination of the commutator structure  below will be the fact that at $x^2=0$ $i\Delta(IT;x^2=0)$ is both singular and independent of $m^2$. As noted in \cite{Schwinger1949} this is due to the fact that there is a discontinuity in going from the mass-dependent timelike $i\Delta(IT;x^2>0)$ to the mass-independent spacelike $i\Delta(IT;x^2<0)$, where because of microcausality the commutator has to vanish. That the mass dependence has to disappear on the light cone is due to the fact that canonical commutators themselves are mass independent (even in a theory with massive fields) and canonical commutators only take support on the light cone.

To establish (\ref{2.5a}) and its light-front analog and identify the relevant singularity structure we need to rewrite (\ref{2.2a}) and (\ref{2.4a}) as four-dimensional Feynman contours: 
\begin{align}
i\Delta(IT;x)&=-\frac{1}{2\pi i}\frac{1}{8\pi^3}\int_{-\infty }^{\infty}dp_1\int_{-\infty }^{\infty}dp_2\int_{-\infty}^{\infty}dp_3\oint dp_0
\nonumber\\
&\times\bigg{[}
\frac{\theta(x^0)e^{-ip\cdot x}-\theta(-x^0)e^{ip\cdot x}}{(p_0)^2-(p_3)^2-(p_1)^2-(p_2)^2-m^2+i\epsilon}
+\frac{\theta(x^0)e^{ip\cdot x}-\theta(-x^0)e^{-ip\cdot x}}{(p_0)^2-(p_3)^2-(p_1)^2-(p_2)^2-m^2-i\epsilon}\bigg{]},
\nonumber\\
i\Delta(LF;x)&=-\frac{1}{2\pi i}\frac{1}{4\pi^3}\int_{-\infty }^{\infty}dp_1\int_{-\infty }^{\infty}dp_2\int_{0}^{\infty}dp_-\oint dp_+
\nonumber\\
&\times\bigg{[}
\frac{\theta(x^+)e^{-ip\cdot x}-\theta(-x^+)e^{ip\cdot x}}{4p_+p_--(p_1)^2-(p_2)^2-m^2+i\epsilon}
+\frac{\theta(x^+)e^{ip\cdot x}-\theta(-x^+)e^{-ip\cdot x}}{4p_+p_--(p_1)^2-(p_2)^2-m^2-i\epsilon}\bigg{]},
\label{2.6a}
\end{align}
with the $+i\epsilon$ terms being closed in the lower-half complex $p_0$ or $p_+$ plane and the $-i\epsilon$ terms being closed in the upper-half plane, and with the signs of $x^0$ and $x^+$ determining whether it is the lower-half circle or the upper-half circle in the complex $p_0$ or $p_+$ plane that is suppressed. We note that in (\ref{2.6a}) $p_-$ is only integrated from $0$ to $\infty$. On rewriting the denominators in (\ref{2.6a}) through the use of the alpha regulators  $\int d\alpha e^{i\alpha(A+i\epsilon)}=-1/i(A+i\epsilon)$ and $\int d\alpha e^{-i\alpha(A-i\epsilon)}=1/i(A-i\epsilon)$ with the $i\epsilon$ suppressing the $\alpha=\infty$ contribution, $\Delta(IT;x)$ and $\Delta(LF;x)$ can then be evaluated as ordinary integrals rather than as contour integrals that have pole terms with singular $1/4p_-$ residues in the  light-front case. The full derivation is given in \cite{Mannheim2019a} and yields  
\begin{align}
i\Delta(IT;x)&=-\frac{1}{2\pi i}\frac{1}{8\pi^3}\int_{-\infty }^{\infty}dp_1\int_{-\infty }^{\infty}dp_2\int_{-\infty}^{\infty}dp_3\int_{-\infty }^{\infty}dp_0\int_0^{\infty}d\alpha\epsilon(x^0)
\nonumber\\
&\times\bigg{[}
-ie^{-ip\cdot x}e^{i\alpha[(p_0)^2-(p_3)^2-(p_1)^2-(p_2)^2-m^2+i\epsilon]}
+ie^{ip\cdot x}e^{-i\alpha[(p_0)^2-(p_3)^2-(p_1)^2-(p_2)^2-m^2-i\epsilon]}\bigg{]}
\nonumber\\
&=-\frac{i}{4\pi^2}\epsilon(x^0)\int_0^{\infty}\frac{d\alpha}{4\alpha^2}\left[e^{-ix^2/4\alpha -i\alpha m^2-\alpha\epsilon}
+e^{ix^2/4\alpha+i\alpha m^2-\alpha \epsilon}\right]
\nonumber\\
&=-\frac{i}{4\pi^2}\epsilon(x^0)\int_0^{\infty}d\beta\left[e^{-i\beta x^2-im^2/4\beta-\beta\epsilon}
+e^{i\beta x^2+im^2/4\beta-\beta\epsilon}\right],
\nonumber\\
i\Delta(LF;x)&=-\frac{1}{2\pi i}\frac{1}{4\pi^3}\int_{-\infty }^{\infty}dp_1\int_{-\infty }^{\infty}dp_2\int_{0}^{\infty}dp_-\int_{-\infty }^{\infty}dp_+\int_0^{\infty}d\alpha
\nonumber\\
&\times\bigg{[}
-i[\theta(x^+)e^{-ip\cdot x}-\theta(-x^+)e^{ip\cdot x}]e^{i\alpha[4p_+p_--(p_1)^2-(p_2)^2-m^2+i\epsilon]}
\nonumber\\
&+i[\theta(x^+)e^{ip\cdot x}-\theta(-x^+)e^{-ip\cdot x}]e^{-i\alpha[4p_+p_--(p_1)^2-(p_2)^2-m^2-i\epsilon]}\bigg{]}
\nonumber\\
&=-\frac{i}{4\pi^2}\epsilon(x^+)\int_0^{\infty}\frac{d\alpha}{4\alpha^2}\left[e^{-ix^2/4\alpha -i\alpha m^2-\alpha\epsilon}
+e^{ix^2/4\alpha+i\alpha m^2-\alpha \epsilon}\right]
\nonumber\\
&=-\frac{i}{4\pi^2}\epsilon(x^+)\int_0^{\infty}d\beta\left[e^{-i\beta x^2-im^2/4\beta-\beta\epsilon}
+e^{i\beta x^2+im^2/4\beta-\beta\epsilon}\right].
\label{2.7a}
\end{align}
In (\ref{2.7a}) we have set $\beta=1/4\alpha$, something that will prove convenient in the following. In (\ref{2.7a}) we not only note the presence of the $\epsilon(x^0)$ term $i\Delta(IT;x)$ that is required in (\ref{2.5a}), we also note that $i\Delta(LF;x)$ has an $\epsilon(x^+)$ counterpart. Since the role of the $i\epsilon$ terms is to indicate how to close the contours in (\ref{2.6a}) we have replaced $-\alpha\epsilon=-\epsilon/4\beta$ by $-\beta\epsilon$ since $\beta$ is positive everywhere in the integration range. 

From  (\ref{2.7a}) we see that $i\Delta(IT;x)$ and $i\Delta(LF;x)$ are extremely similar, and not just similar in fact but actually one and the same function, since under the substitution  $x^0= (x^++x^-)/2$, $x^3= (x^+-x^-)/2$, viz. $x^2=(x^0)^2-(x^3)^2-(x^1)^2-(x^2)^2\rightarrow x^+x^--(x^1)^2-(x^2)^2$ we find that $i\Delta(IT;x)\rightarrow i\Delta(LF;x)$. We have thus achieved our main objective, showing that $i\Delta(IT;x)$ and $ i\Delta(LF;x)$ are related by a coordinate transformation, being in fact related for arbitrary $x^2$, i.e., for timelike, lightlike or spacelike $x^2$. The quantities $i\Delta(IT;x)$ and $ i\Delta(LF;x)$ thus describe the same theory, and we thereby establish that for scalar fields light-front quantization is instant-time quantization. 

As constructed, we see that by rewriting (\ref{2.4a}) as a contour integral and putting the dependence of $p_-$ into the exponentials in the $\alpha$ integrals we no longer have to deal with any $p_-=0$ singularity issues. Moreover, the integrals that appear in (\ref{2.7a})  are standard integrals. For timelike $x^2$ we have
\begin{align}
\theta(x^2)\int _0^{\infty}\frac{d\alpha}{\alpha^2}e^{ix^2/4\alpha+i\alpha m^2-\alpha \epsilon}&=-\theta(x^2)\frac{2m\pi }{(x^2)^{1/2}} [J_1(m(x^2)^{1/2})+iY_1(m(x^2)^{1/2})],
\label{2.8a}
\end{align}
with $i\Delta(IT;x^2>0)$ as given in (\ref{2.5a}) and its $i\Delta(LF;x^2>0)$ analog with $x^0$ replaced by $x^+$, viz.  
\begin{eqnarray}
i\Delta(LF;(x^2> 0)&=&\frac{i}{4\pi}\epsilon(x^+)\frac{mJ_1[m(x^2)^{1/2}]}{(x^2)^{1/2}},
\label{2.9a}
\end{eqnarray}
then following. For spacelike separations $i\Delta(IT;x^2<0)$ and $ i\Delta(LF;x^2<0)$ both vanish. (Since $i\Delta(IT;x^2<0)$ and $ i\Delta(LF;x^2<0)$ vanish when we set  $x^0=0$ or $x^+=0$ in (\ref{2.6a}), by Lorentz invariance they vanish for all spacelike separated distances.)

To treat the $x^2=0$ case in (\ref{2.7a}) we find that at $x^2=0$ both $i\Delta(IT;x^2=0)$ and $ i\Delta(LF;x^2=0)$ diverge, just as is characteristic of the $x^2=0$  delta function that is present in (\ref{2.5a}). However, according to (\ref{2.5a}), this divergence has to be independent of $m^2$. That this is the case is most readily seen in the $\alpha$ form for the integrals that appear in (\ref{2.7a}) as the leading divergence behaves as $\int d\alpha/\alpha^2$ near $\alpha =0$. Since the leading divergence  is independent of $m^2$, we can set $m^2=0$ in the $\beta$ form for the integrals, to then find that at $x^2=0$ we can integrate the $\beta$ integrals directly, to obtain the singular functions
\begin{align}
i\Delta(IT;x^2=0)&=-\frac{i}{4\pi^2}\epsilon(x^0)\left[-\frac{i}{x^2-i\epsilon}+\frac{i}{x^2+i\epsilon}\right]
=-\frac{i}{2\pi}\epsilon(x^0)\delta[(x^0)^2-(x^3)^2-(x^1)^2-(x^2)^2],
\nonumber\\
i\Delta(LF;x^2=0)&=-\frac{i}{2\pi}\epsilon(x^+)\delta[x^+x^--(x^1)^2-(x^2)^2],
\label{2.10a}
\end{align}
with the principal part of $1/(x^2\pm i\epsilon)=PP[1/x^2]\mp i\pi \delta(x^2)$ dropping out. With (\ref{2.10a}) we recover the instant-time $i\Delta(IT;x^2=0)$ given in (\ref{2.5a}), and extend it to $i\Delta(LF;x^2=0)$. With the $\delta[x^+x^--(x^1)^2-(x^2)^2]$ term requiring that $x^+$ and $x^-$ have the same sign, we see that on the light cone $x^0$
has the same sign as $x^+$. Thus in (\ref{2.10a}) we can interchange the $\epsilon(x^0)$ and $\epsilon(x^+)$ terms and thus establish the equivalence of $\Delta(IT;x^2=0)$ and $\Delta(LF;x^2=0)$ all over the light cone.

To check that we have not made a mistake we note that from $i\Delta(IT;x^2=0)$ we can obtain the instant-time (\ref{1.2a}) and (\ref{1.6a}), while from $i\Delta(LF;x^2=0)$ we can obtain the light-front (\ref{1.4a}) and (\ref{1.5a}). Finally, to show the equivalence of  instant-time quantization and light-front time quantization directly starting from $i\Delta(IT;x^2=0)$, we note that at $x^+=0$ the quantity $x^2$ would be negative unless $x^1=0$, $x^2=0$. However, for $x^2<0$ the $i\Delta(IT;x^2<0)$ commutator vanishes. Thus at $x^+=0$ the only point of relevance is $x^2=0$. Now before setting $x^+=0$, in light-front coordinates $i\Delta(IT;x^2=0)$ takes the form
\begin{eqnarray}
i\Delta(IT;x^2=0)=-\frac{i}{2\pi}\epsilon\left[\frac{x^++x^-}{2}\right]\delta[x^+x^--(x^1)^2-(x^2)^2].
\label{2.11a}
\end{eqnarray}
Since $\epsilon(x/2)=\epsilon(x)$ for any $x$, at $x^+=0$ (\ref{2.11a}) takes the form
\begin{eqnarray}
i\Delta(IT;x^2=0)\big{|}_{x^+=0}=-\frac{i}{2\pi}\epsilon(x^-)\delta[(x^1)^2+(x^2)^2].
\label{2.12a}
\end{eqnarray}
Then since $\delta(a^2+b^2)=\pi \delta(a)\delta(b)/2$ for any $a$ and $b$, we can rewrite (\ref{2.12a}) as 
\begin{eqnarray}
i\Delta(IT;x^2=0)\big{|}_{x^+=0}&=&
=-\frac{i}{4}\epsilon(x^-)\delta(x^1)\delta(x^2).
\label{2.13a}
\end{eqnarray}
We recognize (\ref{2.13a}) as (\ref{1.5a}), with the equal light-front time commutation relation (\ref{1.5a}) thus being derived starting from the unequal instant-time commutation relation (\ref{2.2a}) and its complex (\ref{2.6a}) extension. Since the unequal instant-time commutation relation (\ref{2.2a}) itself follows from the equal instant-time commutation relation (\ref{1.2a}) and the scalar field wave equation, we see that the equal light-front time commutation relation (\ref{1.5a}) follows directly from the equal instant-time commutation relation (\ref{1.2a}) (and vice versa) and does not need to be independently postulated. 
Since canonical commutators only take support on the light cone, the equivalence of instant-time quantization and light-front quantization in the scalar field sector is thus established, with the seeming differences between instant-time and light-front commutators only arising because of the restriction of unequal-time commutators to equal times.

\section{Unequal-Time Fermion Field Anticommutators}
\label{S3}

For a fermion field that obeys the free Dirac equation $(i\gamma^{\mu}\partial_{\mu}-m)\psi=0$, the on-shell Fock space expansion of an instant-time fermion field is of the form (see e.g. \cite{Bjorken1965})
\begin{eqnarray}
\psi(x^0,\vec{x})=\sum_{s=\pm}\int \frac{d^3p}{(2\pi)^{3/2}}\left(\frac{m}{E_p}\right)^{1/2}[b(\vec{p},s)u(\vec{p},s)e^{-ip\cdot x}+d^{\dagger}(\vec{p})v(\vec{p},s)e^{+ip\cdot x}],
\label{3.1a}
\end{eqnarray}
where $E_p=+[(p_1)^2+(p_2)^2+(p_3)^2]^{1/2}$, where $s$ denotes the spin projection, where the Dirac spinors $u(\vec{p},s)$ and $v(\vec{p},s)$ obey $(\slashed{p}-m)u(\vec{p},s)=0$, $(\slashed{p}+m)v(\vec{p},s)=0$, and where the non-trivial creation and annihilation operator anticommutation relations are of the form 
\begin{eqnarray}
\{b(\vec{p},s),b^{\dagger}(\vec{q},s^{\prime})\}=\delta_{s,s^{\prime}}\delta^3(\vec{p}-\vec{q}),\quad
\{d(\vec{p},s),d^{\dagger}(\vec{q},s^{\prime})\}=\delta_{s,s^{\prime}}\delta^3(\vec{p}-\vec{q}).
\label{3.2a}
\end{eqnarray}
With these relations  the unequal time anticommutator is given by (see e.g. \cite{Bjorken1965})
\begin{eqnarray}
\big{\{}\psi_{\alpha}(x^0,x^1,x^2,x^3), \psi_{\beta}^{\dagger}(y^0,y^1,y^2,y^3)\big{\}}=\left[(i\gamma^{\mu}\partial_{\mu}+m)\gamma^0\right]_{\alpha\beta}i\Delta(IT;(x-y)^2),
\label{3.3a}
\end{eqnarray}
where $\Delta(IT;(x-y)^2)$ is given in (\ref{2.2a}). From (\ref{3.3a}) the equal instant-time anticommutation relations given in (\ref{1.8a}) then follow.

For the light-front case given that only the good fermion is dynamical, initially it is suggested to generalize the equal light-front time good fermion anticommutator given in (\ref{1.14a}) to unequal light-front time. However, while we might then be able to derive (\ref{1.14a}) starting from (\ref{3.3a}) (which we in fact can \cite{Mannheim2019a,Mannheim2020a}), starting from a projected light-front relation we could not go the other way and derive an instant-time relation from it precisely because projectors are not invertible. However, since $\Lambda^++\Lambda^-=I$, it is only together that the good and bad fermion sectors form a complete basis. Thus to derive instant-time anticommutators from light-front ones, we must start on the fermion light-front side with something that contains all four of the components of the fermion field, and which in addition is invertible. To this end we note that while we could proceed via a Fock space expansion just as we did in the scalar field case, there is actually a more direct procedure. We simply suggest that the light-front analog of (\ref{3.3a}) be given by  
\begin{eqnarray}
\big{\{}\psi_{\alpha}(x^+,x^1,x^2,x^-), \psi_{\beta}^{\dagger}(y^+,y^1,y^2,y^-)\big{\}}=\left[i(\gamma^+\partial_++\gamma^-\partial_-+\gamma^1\partial_1+\gamma^2\partial_2)\gamma^0+m\gamma^0\right]_{\alpha\beta}i\Delta(LF;(x-y)^2),
\label{3.4a}
\end{eqnarray}
and then test for whether or not this might in fact be the case.
We note that in (\ref{3.4a}) we have not transformed $\gamma^0$ into $\gamma^+$, since in going from the instant-time (\ref{1.9a})  to the light-front (\ref{1.10a})  we only transformed the coordinates and not the Dirac gamma matrices 

To establish the validity of (\ref{3.4a}) we note that in since $\psi(x)$ itself obeys the Dirac equation, so does $\big{\{}\psi_{\alpha}(x), \psi_{\beta}^{\dagger}(y)\big{\}}$. However, the Dirac equation is a first-order equation in $\partial/\partial x^+$, so (\ref{3.4a}) will be valid at all $x^+$ if it is valid at one value of $x^+$, which here we take to be $x^+=0$. To check if it is valid at $x^+=0$ we apply $\Lambda^+$ to both sides of (\ref{3.4a}) and also apply $\Lambda^-$ to both sides of (\ref{3.4a}). This yields
\begin{eqnarray}
&&\big{\{}[\psi_{(+)}]_{\alpha}(x^+,x^1,x^2,x^-),[\psi_{(+)}^{\dagger}]_{\beta}(y^+,y^1,y^2,y^-)\big{\}}=2\Lambda^+_{\alpha\beta}i\frac{\partial}{\partial x^-}i\Delta(LF;(x-y)^2),
\label{3.5a}
\end{eqnarray}
\begin{eqnarray}
&&\big{\{}[\psi_{(-)}]_{\alpha}(x^+,x^1,x^2,x^-),[\psi_{(-)}^{\dagger}]_{\beta}(y^+,y^1,y^2,y^-)\big{\}}=2\Lambda^-_{\alpha\beta}i\frac{\partial}{\partial x^+}i\Delta(LF;(x-y)^2).
\label{3.6a}
\end{eqnarray}
On the light cone $i\Delta(LF;(x-y)^2)$ is given by (\ref{2.10a}), and in it we can replace $\epsilon(x^+-y^+)$ by $\epsilon(x^--y^-)$ since the delta function in (\ref{2.10a}) requires that $x^+-y^+$ and $x^--y^-$ have the same sign. Thus from (\ref{3.5a})  we obtain 
\begin{align}
&\big{\{}[\psi_{(+)}]_{\alpha}(x^+,x^1,x^2,x^-),[\psi_{(+)}^{\dagger}]_{\beta}(y^+,y^1,y^2,y^-)\big{\}}|_{(x-y)^2=0}
\nonumber\\
&=2i\Lambda^+_{\alpha\beta}\left(\frac{-i}{2\pi}\right)\bigg{[}2\delta(x^--y^-)
\delta[(x^+-y^+)(x^--y^-)-(x^1-y^1)^2-(x^2-y^2)^2]
\nonumber\\
&+(x^+-y^+)\epsilon(x^--y^-)
\delta^{\prime}[(x^+-y^+)(x^--y^-)-(x^1-y^1)^2-(x^2-y^2)^2
\bigg{]}
\label{3.7a}
\end{align}
in the good fermion sector.
At $x^+=y^+$ we obtain
\begin{eqnarray}
&&\big{\{}[\psi_{(+)}]_{\alpha}(x^+,x^1,x^2,x^-),[\psi_{(+)}^{\dagger}]_{\beta}(x^+,y^1,y^2,y^-)\big{\}}
=\Lambda^+_{\alpha\beta}\delta(x^--y^-)\delta(x^1-y^1)\delta(x^2-y^2).
\label{3.8a}
\end{eqnarray}
We thus obtain the good fermion (\ref{1.14a}).

In order to evaluate the right-hand side of (\ref{3.6a}) in the bad fermion case we have found it convenient to use the form for $i\Delta(LF;(x-y)^2)$ given in (\ref{2.6a}). On applying $\partial/\partial x^+$ to $i\Delta(LF;(x-y)^2)$ we obtain
\begin{align}
&\frac{\partial}{\partial x^+}i\Delta(LF;(x-y)^2)=-\frac{1}{2\pi i}\frac{1}{4\pi^3}\int_{-\infty }^{\infty}dp_1\int_{-\infty }^{\infty}dp_2\int_{0}^{\infty}dp_-\int_{-\infty }^{\infty}dp_+
\nonumber\\
&\times\bigg{[}
\delta(x^+-y^+)\frac{e^{-ip\cdot (x-y)}+e^{ip\cdot (x-y)}}{4p_+p_--(p_1)^2-(p_2)^2-m^2+i\epsilon}
+\delta(x^+-y^+)\frac{e^{ip\cdot (x-y)}+e^{-ip\cdot (x-y)}}{4p_+p_--(p_1)^2-(p_2)^2-m^2-i\epsilon}
\nonumber\\
&+\frac{\theta(x^+-y^+)(-ip_+)e^{-ip\cdot (x-y)}-\theta(-x^++y^+)(ip_+)e^{ip\cdot (x-y)}}{4p_+p_--(p_1)^2-(p_2)^2-m^2+i\epsilon}
\nonumber\\
&+\frac{\theta(x^+-y^+)(ip_+)e^{ip\cdot (x-y)}-\theta(-x^++y^+)(-ip_+)e^{-ip\cdot (x-y)}}{4p_+p_--(p_1)^2-(p_2)^2-m^2-i\epsilon}\bigg{]}.
\label{3.9a}
\end{align}
In each of the terms that contain a delta function the delta functions cause all the $\pm ip_+(x^+-y^+)$ terms in the exponents to vanish identically. However that then causes the residues at the poles in the $+i\epsilon$ and $-i\epsilon$ terms to be equal. Then since the $+i\epsilon$ term is closed below the real $p_+$ axis in a clockwise direction while the $-i\epsilon$ term is closed above the real $p_+$ axis in a counter-clockwise direction the two delta function terms cancel each other identically.

In order to make contact with (\ref{1.18a}), from which (\ref{1.19a}) follows, we apply $\partial/\partial y^-$ to the 
theta-function-dependent terms in (\ref{3.9a}). This yields
\begin{align}
&\frac{\partial}{\partial y^-}\frac{\partial}{\partial x^+}i\Delta(LF;(x-y)^2)=-\frac{1}{2\pi i}\frac{1}{4\pi^3}\int_{-\infty }^{\infty}dp_1\int_{-\infty }^{\infty}dp_2\int_{0}^{\infty}dp_-\int_{-\infty }^{\infty}dp_+
\nonumber\\
&\times\bigg{[}
\frac{\theta(x^+-y^+)p_+p_-e^{-ip\cdot (x-y)}-\theta(-x^++y^+)p_+p_-e^{ip\cdot (x-y)}}{4p_+p_--(p_1)^2-(p_2)^2-m^2+i\epsilon}
\nonumber\\
&+\frac{\theta(x^+-y^+)p_+p_-e^{ip\cdot (x-y)}-\theta(-x^++y^+)p_+p_-e^{-ip\cdot (x-y)}}{4p_+p_--(p_1)^2-(p_2)^2-m^2-i\epsilon}\bigg{]}.
\label{3.10a}
\end{align}
Since the only contributions to the contour integrals are poles we can replace the $p_+p_-$ terms in the numerators by $[(p_1)^2+(p_2)^2+m^2]/4$. We can then replace these terms by derivatives with respect to $x^1$ and $x^2$, and obtain
\begin{align}
&\frac{\partial}{\partial y^-}\frac{\partial}{\partial x^+}i\Delta(LF;(x-y)^2)=-\frac{1}{2\pi i}\frac{1}{4\pi^3}\int_{-\infty }^{\infty}dp_1\int_{-\infty }^{\infty}dp_2\int_{0}^{\infty}dp_-\int_{-\infty }^{\infty}dp_+\frac{1}{4}\left[-\frac{\partial}{\partial x^1}\frac{\partial}{\partial x^1}-\frac{\partial}{\partial x^2}\frac{\partial}{\partial x^2}+m^2\right]
\nonumber\\
&\times\bigg{[}
\frac{\theta(x^+-y^+)e^{-ip\cdot (x-y)}-\theta(-x^++y^+)e^{ip\cdot (x-y)}}{4p_+p_--(p_1)^2-(p_2)^2-m^2+i\epsilon}
\nonumber\\
&+\frac{\theta(x^+-y^+)e^{ip\cdot (x-y)}-\theta(-x^++y^+)e^{-ip\cdot (x-y)}}{4p_+p_--(p_1)^2-(p_2)^2-m^2-i\epsilon}\bigg{]}.
\label{3.11a}
\end{align}
Comparing with (\ref{2.6a}) we thus obtain 
\begin{align}
&\frac{\partial}{\partial y^-}\frac{\partial}{\partial x^+}i\Delta(LF;(x-y)^2)=\frac{1}{4}\left[-\frac{\partial}{\partial x^1}\frac{\partial}{\partial x^1}-\frac{\partial}{\partial x^2}\frac{\partial}{\partial x^2}+m^2\right]i\Delta(LF;(x-y)^2).
\label{3.12a}
\end{align}
Finally, on taking a $\partial/\partial x^-$ derivative and applying the $\Lambda^-$ projection operator to (\ref{3.12a}), from (\ref{3.6a}) we obtain 
\begin{align}
&\Big{\{}\frac{\partial}{\partial x^-}\psi_{\alpha}^{(-)}(x^+,x^1,x^2,x^-),\frac{\partial}{\partial y^-}[\psi_{(-)}^{\dagger}]_{\beta}(y^+,y^1,y^2,y^-)\Big{\}}
\nonumber\\
&=2i\Lambda^-_{\alpha\beta}\frac{1}{4}\left[-\frac{\partial}{\partial x^1}\frac{\partial}{\partial x^1}-\frac{\partial}{\partial x^2}\frac{\partial}{\partial x^2}+m^2\right]\frac{\partial}{\partial x^-}i\Delta(LF;(x-y)^2).
\label{3.13a}
\end{align}

At $x^+-y^+=0$ the quantity $(x-y)^2$ could only be lightlike or spacelike. But $i\Delta(LF;(x-y)^2)$ vanishes for spacelike separations, so at $x^+=y^+$ the quantity $(x-y)^2$ must be zero. With $i\Delta(LF;(x-y)^2=0)$ being given in  (\ref{2.10a}), and with (\ref{2.10a}) being rewritable as 
$i\Delta(LF;(x-y)^2=0)=-(i/2\pi)\epsilon(x^--y^-)\delta[(x^+-y^+)(x^--y^-)-(x^1-y^1)^2-(x^2-y^2)^2]$ since the delta function forces $x^+-y^+$ and $x^--y^-$ to have the same sign, on the light cone the equal light-front time bad fermion anticommutator evaluates to
\begin{align}
&\Big{\{}\frac{\partial}{\partial x^-}\psi_{\alpha}^{(-)}(x^+,x^1,x^2,x^-),\frac{\partial}{\partial y^-}[\psi_{(-)}^{\dagger}]_{\beta}(x^+,y^1,y^2,y^-)\Big{\}}
\nonumber\\
&=2i\Lambda^-_{\alpha\beta}\frac{1}{4}\left[-\frac{\partial}{\partial x^1}\frac{\partial}{\partial x^1}-\frac{\partial}{\partial x^2}\frac{\partial}{\partial x^2}+m^2\right]\frac{\partial}{\partial x^-}\left[-\frac{i}{2\pi}\epsilon(x^--y^-)\frac{\pi}{2}\delta(x^1-y^1)\delta(x^2-y^2)\right]
\nonumber\\
&=\frac{1}{4}\Lambda^-_{\alpha\beta}\left[-\frac{\partial}{\partial x^1}\frac{\partial}{\partial x^1}-\frac{\partial}{\partial x^2}\frac{\partial}{\partial x^2}+m^2\right]\delta(x^--y^-)\delta(x^1-y^1)\delta(x^2-y^2).
\label{3.14a}
\end{align}
We recognize (\ref{3.14a}) as (\ref{1.18a}). By the same procedure we can also recover (\ref{1.20a}) \cite{Mannheim2019a}, and thus can recover the full set of good fermion good fermion, good fermion bad fermion, and bad fermion bad fermion anticommutators. With the good and bad fermions together being complete since $\Lambda^++\Lambda^-=I$, we thus confirm that the expression for the light-front $\big{\{}\psi_{\alpha}(x^+,x^1,x^2,x^-), \psi_{\beta}^{\dagger}(y^+,y^1,y^2,y^-)\big{\}}$ given in (\ref{3.4a}) is indeed valid. Comparing (\ref{3.4a}) with the instant-time $\big{\{}\psi_{\alpha}(x^0,x^1,x^2,x^3), \psi_{\beta}^{\dagger}(y^0,y^1,y^2,y^3)\big{\}}$ given in (\ref{3.3a}), we see that the discussion can now completely parallel the scalar field case. We thus establish that even with non-invertible projection operators, just as in the scalar field case, equally for fermions light-front quantization is instant-time quantization.

\section{Equivalence for Abelian Gauge Fields}
\label{S4}

For our purposes here rather than choose a gauge as is common in gauge field studies, we have found it to be more  convenient to use Feynman gauge fixing. Thus we take the gauge field action $I_G$ in both the instant-time and light-front cases to be of the gauge fixing form 
\begin{eqnarray}
I_G=\int d^4x\left[ -\tfrac{1}{4}F_{\mu\nu}F^{\mu\nu}-\tfrac{1}{2}(\partial_{\mu}A^{\mu})^2\right],
\label{4.1}
\end{eqnarray}
where $A_{\mu}$ is an Abelian gauge field and $F_{\mu\nu}=\partial_{\nu}A_{\mu}-\partial_{\mu}A_{\nu}$. Since surface terms are held fixed in a functional variation, they make no contribution to the field equations that are obtained by stationary variation of the action. Thus on integrating $I_G$ by parts and dropping surface terms the action simplifies to
\begin{eqnarray}
I_G=\int d^4x\left[ -\tfrac{1}{2}\partial_{\nu}A_{\mu}\partial^{\nu}A^{\mu}\right].
\label{4.2}
\end{eqnarray}
Variation of the $I_G$ action with respect to $A_{\mu}$ yields an equation of motion of the form
\begin{eqnarray}
\partial_{\nu}\partial^{\nu}A_{\mu}=0.
\label{4.3}
\end{eqnarray}
The utility of using (\ref{4.2}) is that the various components of $A_{\mu}$ are decoupled from each other in the equation of motion.  Consequently, we can treat each component of $A_{\mu}$ as an independent degree of freedom, and apply the scalar field analysis given above to each one of them. In this formulation (\ref{4.3}) entails that $\partial_{\nu}\partial^{\nu}\chi=0$, where $\chi=\partial_{\mu}A^{\mu}$. If one imposes the subsidiary conditions $\chi(\tau=0)=0$, $\partial_0\chi(\tau=0)=0$ at the initial time $\tau=0$ (with $\tau=x^0$ or $\tau=x^+$), then since $\partial_{\nu}\partial^{\nu}\chi=0$ is a second-order derivative equation it follows that the non-gauge-invariant $\chi$ is zero at all times.

Given (\ref{4.2}) one can define instant-time canonical conjugates of the form $\Pi^{\mu}=\delta I_{G}/\delta \partial_0A_{\mu}=-\partial^0A^{\mu}$. This then leads to equal instant-time commutation relations of the form (see e.g. \cite{Schweber1964})
\begin{align}
&[A_{\nu},\Pi^{\mu}]=[A_{\nu}(x^0,x^1,x^2,x^3),-\partial^0A^{\mu}(x^0,y^1,y^2,y^3)]=-i\delta^{\mu}_{\nu}\delta(x^1-y^1)\delta(x^2-y^2)\delta(x^3-y^3),
\nonumber\\
&[A_{\nu}(x^0,x^1,x^2,x^3),\partial_0A_{\mu}(x^0,y^1,y^2,y^3)]=ig_{\mu\nu}\delta(x^1-y^1)\delta(x^2-y^2)\delta(x^3-y^3),
\label{4.4}
\end{align}
and in analog to the scalar field case, to unequal instant-time commutation relations of the form (see e.g. \cite{Schweber1964})
\begin{align}
&[A_{\nu}(x^0,x^1,x^2,x^3),A_{\mu}(y^0,y^1,y^2,y^3)]=ig_{\mu\nu}\Delta(IT;(x-y)^2)
\nonumber\\
&=-\frac{i}{2\pi}g_{\mu\nu}\epsilon(x^0-y^0)\delta[(x^0)^2-(x^1)^2-(x^2)^2-(x^3)^2],
\label{4.5}
\end{align}
where $g_{\mu\nu}$ is the instant-time metric and $i\Delta(IT;(x-y)^2)$ is the scalar field $i\Delta(IT;(x-y)^2)$ as given in (\ref{2.6a}) when we set $m=0$ (gauge fields being massless). Since $i\Delta(IT;((x-y)^2>0)$ as given in (\ref{2.5a}) vanishes when $m=0$, and since $i\Delta(IT;((x-y)^2<0)$ as given in  \ref{2.5a}) is zero for any $m$, it follows that when $m=0$ $i\Delta(IT;(x-y)^2)$ is given by $i\Delta(IT;((x-y)^2=0)$ as given in  (\ref{2.5a}).

Given (\ref{4.2}) one can also define equal light-front time canonical conjugates of the form $\Pi^{\mu}=\delta I_{G}/\delta \partial_+A_{\mu}=-\partial^+A^{\mu}=-2\partial_-A^{\mu}$. This leads to equal light-front time commutation relations  of a form analogous to (\ref{1.4a}) and (\ref{1.5a}), viz. \cite{Mannheim2019a}
\begin{align}
&[A_{\nu},\Pi^{\mu}]=[A_{\nu}(x^+,x^1,x^2,x^-), -2\partial_-A^{\mu}(x^+,y^1,y^2,y^-)]=-i\delta_{\mu}^{\nu}\delta(x^1-y^1)\delta(x^2-y^2)\delta(x^--y^-),
\nonumber\\
&[A_{\nu}(x^+,x^1,x^2,x^-),\partial_-A_{\mu}(x^+,y^1,y^2,y^-)]=\frac{i}{2}g_{\mu\nu}\delta(x^1-y^1)\delta(x^2-y^2)\delta(x^--y^-),
\nonumber\\
&[A_{\nu}(x^+,x^1,x^2,x^-), A_{\mu}(x^+,y^1,y^2,y^-)]=-\frac{i}{4}g_{\mu\nu}\epsilon(x^--y^-)\delta(x^1-y^1)\delta(x^2-y^2),
\label{4.6}
\end{align}
where $g_{\mu\nu}$ is the light-front metric. Thus, in analog to the scalar field case, the last expression in the equal light-front time (\ref{4.6})  follows directly from the unequal instant-time (\ref{4.5}), with the instant-time metric transforming into the light-front metric. Thus for Abelian gauge fields we again see that light-front quantization is instant-time quantization. 

\section{Comparing Gauge Fixing with a Choice of Gauge}
\label{S5}

While on the topic of gauge fields, we note that our particular use of gauge fixing leads us to gauge field equations of motion that, because of our having integrated the action by parts, are diagonal in the spacetime indices, to thus be associated with propagators of the form $D^{\mu\nu}(p)=g^{\mu\nu}/(p^2+i\epsilon)$ that are equally diagonal in the spacetime indices. This is to be contrasted with quantization of the unmodified Maxwell action $I_M= -\tfrac{1}{4}\int d^4xF_{\mu\nu}F^{\mu\nu}$ in the $n_{\mu}A^{\mu}=0$ gauge with general fixed spacelike reference vector $n_{\mu}$, a gauge commonly used in light-front gauge field studies. In this gauge the light-front gauge field propagator is conveniently given by \cite{Harindranath1996} 
\begin{eqnarray}
D^{\mu\nu}(x)=-i\langle \Omega|T[A^{\mu}(x)A^{\nu}(0)]|\Omega\rangle=2\int \frac{dp_+dp_-dp_1dp_2}{(2\pi)^4}\frac{e^{-ip\cdot x}}{p^2+i\epsilon}\left(g^{\mu\nu}-\frac{n^{\mu}p^{\nu}+n^{\nu}p^{\mu}}{n\cdot p}
+\frac{p^2}{(n\cdot p)^2}n^{\mu}n^{\nu}\right),
\label{5.1}
\end{eqnarray}
to thus not only not be diagonal in the spacetime indices, but, as had been noted in  \cite{Yan1973,Leibbrandt1984}, to also contain terms that are singular at $n\cdot p=0$. We note that unlike commutators such as $i\Delta(LF;x)$ where $p_-$ ranges between $0$ and $\infty$, in propagators  such as the one in (\ref{5.1}) $p_-$ ranges between $-\infty$ and $\infty$ as is needed since the wave operator acting on a propagator has to generate a four-dimensional delta function of the form  $2\delta(x^+)\delta(x^1)\delta(x^2)\delta(x^-)$.

Since none of the singular terms that are present in (\ref{5.1}) appear in propagators constructed with our gauge fixed $I_G$, the origin of the $n\cdot p$ dependent terms in (\ref{5.1}) is due to using $I_M$ and making a gauge choice. For instance when one takes the gauge to be $A^+=0$, i.e., $n_+=1$, $n^-=2$ (all other components of $n^{\mu}$ and $n_{\mu}$ zero), then while $D^{11}$ and $D^{22}$ only contain the $g^{\mu\nu}$ term, $D^{-1}$ and $D^{-2}$ contain one and $D^{--}$ contains both of the two terms in (\ref{5.1}) that are singular at $p^+=2p_-=0$. As noted in \cite{Mannheim2019a}, in the $A^+=0$ gauge $D^{11}$ and $D^{22}$ obey
\begin{eqnarray}
&&[\eta_{1\nu}\partial_{\alpha}\partial^{\alpha}-\partial_{1}\partial_{\nu}]D^{\nu }_{\phantom{\nu}1}
=[\eta_{2\nu}\partial_{\alpha}\partial^{\alpha}-\partial_{2}\partial_{\nu}]D^{\nu}_{\phantom{\nu} 2}
=2\delta(x^+)\delta(x^-)\delta(x^1)\delta(x^2).
\label{5.2}
\end{eqnarray}
On inserting (\ref{5.1}) into (\ref{5.2}) we find that the form given in (\ref{5.1}) satisfies (\ref{5.2}) identically, as it of course would have to since both sets of equations are associated with quantizing $I_M$ in the $A^+=0$ gauge. To see this in detail we note that we can break up (\ref{5.2}) into two separate sectors each one of which satisfies (\ref{5.2}) separately. Specifically we find that 
\begin{eqnarray}
-\partial_{1}\partial_{\nu}D^{\nu }_{\phantom{\nu}1}=-\partial_{1}\partial_{1}D^{1}_{\phantom{1}1}-\partial_{1}\partial_{-}D^{- }_{\phantom{-}1}=0,\quad
-\partial_{2}\partial_{\nu}D^{\nu}_{\phantom{\nu} 2}=-\partial_{2}\partial_{2}D^{2}_{\phantom{2}2}-\partial_{2}\partial_{-}D^{- }_{\phantom{-}2}=0,
\label{5.3}
\end{eqnarray}
and are thus left with
\begin{eqnarray}
&&\eta_{11}\partial_{\alpha}\partial^{\alpha}D^{1 }_{\phantom{1}1}=-\partial_{\alpha}\partial^{\alpha}D^{1}_{\phantom{1}1}
=\eta_{22}\partial_{\alpha}\partial^{\alpha} D^{2}_{\phantom{2} 2}=-\partial_{\alpha}\partial^{\alpha}D^{2}_{\phantom{2}2}
=2\delta(x^+)\delta(x^-)\delta(x^1)\delta(x^2),
\label{5.4}
\end{eqnarray}
with all the singular terms cancelling against each other in $D^{11}$ and $D^{22}$ so that $D^{11}(p)=g^{11}/(p^2+i\epsilon)$ and $D^{22}(p)=g^{22}/(p^2+i\epsilon)$. 

In regard to the singular terms that appear in (\ref{5.1}), as noted in \cite{Harindranath1996} these extra terms arise because in the $A^+=0$ gauge $A^-$ is a constrained field that obeys the nonlocal constraint
\begin{eqnarray}
&&2\partial_-(\partial_-A^-+\partial_1A^1+\partial_2A^2)=0,\quad A^-=-\int duD_2(x^--u)\partial_-(\partial_1A^1+\partial_2A^2),
\label{5.5}
\end{eqnarray}
associated with the Maxwell equations $\partial_{\nu}F^{\mu\nu}=0$, where $D_2=(\partial_-)^2$ is an inverse propagator. Specifically, we note that the factor $g^{\mu\nu}-(n^{\mu}p^{\nu}+n^{\nu}p^{\mu})/n\cdot p+p^2n^{\mu}n^{\nu}/(n\cdot p)^2$ on the right-hand side of (\ref{5.1}) evaluates to $-(p_1/p_-)^2-(p_2/p_-)^2$ when $\mu=-$, $\nu=-$, precisely as is needed to verify the validity of (\ref{5.1}) in the $(-,-)$ sector by inserting the nonlocal (\ref{5.5})  into $D^{--}(x)=-i\langle \Omega|T[A^{-}(x)A^{-}(0)]|\Omega\rangle$. With $A^+$, $g^{++}$ and $n^+$ all being zero, (\ref{5.1}) holds for all components of  $D^{\mu\nu}(x)$. And with $A^-$ being constrained, only $A^1$ and $A^2$ propagate, and for them we can use $D^{\mu\nu}(p)=g^{\mu\nu}/(p^2+i\epsilon)$. Singularities do appear in $D^{--}$, $D^{-1}$ and $D^{-2}$, and with (\ref{5.1}) being local in momentum space, $D^{--}$, $D^{-1}$ and $D^{-2}$ are nonlocal in coordinate space, just as $A^-$ is. While this nonlocality is analogous to the bad fermion anticommutator given in (\ref{1.19a}), which is also nonlocal in coordinate space, the two situations are not comparable since the bad fermion obeys a nonlocal constraint due to the intrinsic structure of the light-front Dirac equation ($\gamma^+$ being a divisor of zero). However, $A^+$ only obeys a nonlocal constraint because of the gauge choice, and with $I_G$ leading to $D^{\mu\nu}(p)=g^{\mu\nu}/(p^2+i\epsilon)$ for all $\mu$, $\nu$, the associated singularities in (\ref{5.1}) would even appear to be avoidable since using our gauge fixing action $I_G$ apparently leads to no singularities at all. Thus it would appear that the treatments of these $n\cdot p=0$ singularities by Mandelstam \cite{Mandelstam1983} and Leibbrandt \cite{Leibbrandt1984} might not be needed, because with $I_G$ they do not appear. Also we note that in establishing the equivalence of instant-time and light-front vacuum tadpole graphs given in \cite{Mannheim2019a,Mannheim2019b} it is necessary to deal with the $p^+=2p_-=0$ region, as even in the scalar field case where there are no gauge issues at all, this zero-mode region puts singularities into Feynman diagrams (the $p_+$ pole term in (\ref{7.8a}) below generates an on-shell  $1/4p_-$ term, just like the one in the on-shell (\ref{2.4a})). These zero-mode  Feynman diagram singularities are distinct from those in (\ref{5.1}), and are characteristic of light-front studies. They have been treated quite extensively in   \cite{Mannheim2019a,Mannheim2019b} and will be discussed briefly in Sec. \ref{S8} below.

\section{Equivalence for non-Abelian Gauge Fields}
\label{S6}

In the Yang-Mills case one has a non-Abelian group with structure coefficients $f_{abc}$. One defines a tensor $G^a_{\mu\nu}=\partial_{\nu}A^a_{\mu}-\partial_{\mu}A^a_{\nu}+gf^{abc}A^b_{\nu}A^c_{\mu}$ where $g$ is the coupling constant. In analog to (\ref{4.1}) one defines an action (see e.g. \cite{Donoghue1992}) 
\begin{eqnarray}
I_{YM}=\int d^4x\left[ -\tfrac{1}{4}G^a_{\mu\nu}G_a^{\mu\nu}-\tfrac{1}{2}\partial_{\mu}A_a^{\mu}\partial_{\nu}A_a^{\nu}
+\partial_{\mu}\bar{c}_a\partial^{\mu}c_a+gf^{abe}A^{\mu}_a\partial_{\mu}\bar{c}_bc_e\right],
\label{6.1}
\end{eqnarray}
where the $c_a$ and $\bar{c}_a$ are Fadeev-Popov ghost fields that one has to introduce in the non-Abelian case, viz. spin zero Grassmann fields that are quantized with  anticommutation relations. Since the $g$-dependent terms in $I_{YM}$ involve products of either three or four fields they can be treated as part of the interaction. On integrating by parts, the relevant part of $I_{YM}$ for quantization, viz. the free part,  is thus given by
\begin{eqnarray}
I_{YM}=\int d^4x\left[ -\tfrac{1}{2}\partial_{\nu}A^a_{\mu}\partial^{\nu}A_a^{\mu}+\partial_{\mu}\bar{c}_a\partial^{\mu}c_a\right],
\label{6.2}
\end{eqnarray}
and leads to  equations of motion of the form
\begin{eqnarray}
\partial_{\nu}\partial^{\nu}A_{\mu}^a=0,\quad \partial_{\mu}\partial^{\mu}c_a=0,\quad \partial_{\mu}\partial^{\mu}\bar{c}_a=0.
\label{6.3}
\end{eqnarray}
With both (\ref{6.2}) and (\ref{6.3}) being diagonal in both spacetime and group indices, the discussion thus parallels the Abelian and scalar field cases, with $A_{\mu}^a$ acting the same way as the Abelian $A_{\mu}$ and $c_a$ and $\bar{c}_a$ acting the same way as $\phi$. And in addition, with the perturbative instant-time gauge boson and ghost propagators being of the respective forms $g_{\mu\nu}\delta_{ab}/(p^2+i\epsilon)$ and $\delta_{ab}/(p^2+i\epsilon)$, the perturbative instant-time gauge boson and ghost propagators transform into the perturbative light-front  gauge boson and ghost propagators.  Thus as with the Abelian case, in the non-Abelian case light-front quantization again is instant-time quantization. And moreover, just as in the Abelian case, through our use of gauge fixing no zero-mode singularities appear in the propagators.

\section{Time-Ordered Products of Operators}
\label{S7}

For scalar fields in either instant-time or light-front quantization the propagator that satisfies the wave equation with a delta function source is given by $-i\langle \Omega|T[\phi(x)\phi(0)]|\Omega\rangle$. This is also the case for the fermionic $-i\langle \Omega|T[\psi(x)\bar{\psi}(0)]|\Omega\rangle$ in instant-time quantization as it is given by
\begin{align}
&-i\langle \Omega|[\theta(x^0)\psi_{\beta}(x)\bar{\psi}_{\alpha}(0)-\theta(-x^0)\bar{\psi}_{\alpha}(0)\psi_{\beta}(x)]\Omega \rangle
\nonumber\\
&=\frac{1}{(2\pi)^4}\int dp_0dp_1dp_2dp_3\Big{[}\frac{e^{-i(p_0x^0+p_1x^1+p_2x^2+p_3x^3)}}{\gamma^0p_0+\gamma^3p_3+\gamma^1p_1+\gamma^2p_2-m+i\epsilon}\Big{]}_{\beta\alpha}=S^{IT}_F(x)_{\beta\alpha},
\label{7.1a}
\end{align}
with $S^{IT}_F(x)_{\beta\alpha}$ being the instant-time Feynman propagator. However, it is not the case for fermions in light-front quantization. Specifically, in the fermion light-front case the time-ordered product is given by  \cite{Yan1973}
\begin{align}
&-i\langle \Omega|[\theta(x^+)\psi_{\beta}(x)\bar{\psi}_{\alpha}(0)-\theta(-x^+)\bar{\psi}_{\alpha}(0)\psi_{\beta}(x)]\Omega \rangle
\nonumber\\
&=\frac{2}{(2\pi)^4}\int dp_+dp_1dp_2dp_-\Big{[}\frac{e^{-i(p_+x^++p_1x^1+p_2x^2+p_-x^-)}}{\gamma^+p_++\gamma^-p_-+\gamma^1p_1+\gamma^2p_2-m+i\epsilon}\Big{]}_{\beta\alpha}
+\frac{i}{4}\gamma^+_{\beta\alpha}\delta(x^+)\epsilon(x^-)\delta(x^1)\delta(x^2)
\nonumber\\
&=S^{LF}_F(x)_{\beta\alpha}+\frac{i}{4}\gamma^+_{\beta\alpha}\delta(x^+)\epsilon(x^-)\delta(x^1)\delta(x^2),
\label{7.2a}
\end{align}
with $S^{LF}_F(x)_{\beta\alpha}$ being the light-front Feynman propagator. So again instant-time and light-front quantization appear to be different. 

In keeping with our exploration of operators themselves rather than their matrix elements, we now seek to understand the difference between and then the relationship  between (\ref{7.1a}) and (\ref{7.2a}) from an operator perspective. To this end we have found it convenient to look not at the time-ordered q-number operator $T[\psi_{\beta}(x)\bar{\psi}_{\alpha}(0)]$  itself but rather at its Dirac operator derivative. First, we note that  through use of the instant-time anticommutator given in (\ref{1.8a}), the instant-time q-number time-ordered product obeys
\begin{align}
&[i\gamma^{\mu}\partial_{\mu}-m]_{\lambda\beta}(-i)[\theta(x^0)\psi_{\beta}(x)\bar{\psi}_{\alpha}(0)-\theta(-x^0)\bar{\psi}_{\alpha}(0)\psi_{\beta}(x)]
\nonumber\\
&=i\gamma^0_{\lambda\beta}(-i)\delta(x^0)[\psi_{\beta}(x)\bar{\psi}_{\alpha}(0)+\bar{\psi}_{\alpha}(0)\psi_{\beta}(x)]
=\delta_{\lambda\alpha}\delta(x^0)\delta(x^3)\delta(x^1)\delta(x^2).
\label{7.3a}
\end{align}
Thus we see  that even though the time-ordered product is itself a q-number, its Dirac operator derivative  is a c-number. This is analogous to our discussion of $-i\Delta(IT;x)$ and  $-i\Delta(LF;x)$, with the unequal-time $-i\Delta(IT;x)$ and  $-i\Delta(LF;x)$ also being c-numbers. With time-ordered products also being defined at unequal times, we see that the discussion of the Dirac operator derivative  of the q-number time-ordered product will parallel that of our earlier discussion of $-i\Delta(IT;x)$ and its Dirac operator derivative  given in (\ref{3.3a}), as it has the light-front analog that is given in (\ref{3.4a}).

We thus apply the Dirac operator derivative  to  the fermion time-ordered operator product at unequal light-front time, and obtain  
\begin{align}
&[i\gamma^{\mu}\partial_{\mu}-m]_{\lambda\beta}(-i)[\theta(x^+)\psi_{\beta}(x)\bar{\psi}_{\alpha}(0)-\theta(-x^+)\bar{\psi}_{\alpha}(0)\psi_{\beta}(x)]
\nonumber\\
&=i\gamma^+_{\lambda\beta}(-i)\delta(x^+)[\psi_{\beta}(x)\bar{\psi}_{\alpha}(0)+\bar{\psi}_{\alpha}(0)\psi_{\beta}(x)].
\label{7.4a}
\end{align}
To be able to evaluate this expression using equal light-front time anticommutators we insert $\Lambda^++\Lambda^-=I$, and with $\Lambda^{\pm}=\gamma^0\gamma^{\pm}/2$ obtain 
\begin{align}
&[i\gamma^{\mu}\partial_{\mu}-m]_{\lambda\beta}(-i)[\theta(x^+)\psi_{\beta}(x)\bar{\psi}_{\alpha}(0)-\theta(-x^+)\bar{\psi}_{\alpha}(0)\psi_{\beta}(x)]
\nonumber\\
&=2\gamma^0_{\lambda\nu}\delta(x^+) \Big{\{}[\psi_{(+)}]_{\nu}(x),[\psi^{\dagger}_{(+)}]_{\sigma}(0)\Big{\}}\gamma^0_{\sigma\alpha}
+2\gamma^0_{\lambda\nu}\delta(x^+) \Big{\{}[\psi_{(+)}]_{\nu}(x),[\psi^{\dagger}_{(-)}]_{\sigma}(0)\Big{\}}\gamma^0_{\sigma\alpha}.
\label{7.5a}
\end{align}
Then given (\ref{1.14a}) and (\ref{1.20a}), following some algebra we obtain \cite{Mannheim2019a}
\begin{align}
&[i\gamma^{\mu}\partial_{\mu}-m]_{\lambda\beta}(-i)[\theta(x^+)\psi_{\beta}(x)\bar{\psi}_{\alpha}(0)-\theta(-x^+)\bar{\psi}_{\alpha}(0)\psi_{\beta}(x)]
\nonumber\\
&=2\delta_{\lambda\alpha}\delta(x^+)\delta(x^-)\delta(x^1)\delta(x^2)-\tfrac{1}{2}[\gamma^-\gamma^+]_{\lambda\alpha}\delta(x^+)\delta(x^-)\delta(x^1)\delta(x^2)
\nonumber\\
&+\tfrac{i}{4}\delta(x^+)\epsilon(x^-)[i(\gamma^1\gamma^+\partial_1+\gamma^2\gamma^+\partial_2)-m\gamma^+]_{\lambda\alpha}\delta(x^1)\delta(x^2),
\label{7.6a}
\end{align}
Thus just as in the instant-time case,  the action of the Dirac operator derivative on the q-number light-front time-ordered fermion product yields a c-number. In (\ref{7.6a}) the first delta function term is the light-front analog of the delta function term in the instant-time (\ref{7.3a}). The latter two terms in (\ref{7.6a}) are intrinsic to light-front quantization.  Finally, in order to now go from (\ref{7.6a}) to (\ref{7.2a}) we take the vacuum matrix element of (\ref{7.6a}). This then gives us a pure c-number differential equation, and we find that its solution is none other than (\ref{7.2a}). The singular term in (\ref{7.2a}) is thus associated with the singular terms in (\ref{7.6a}).

Now despite the fact that (\ref{7.3a}) and (\ref{7.6a}) look to be quite different from each other, we note that by a coordinate transformation we can transform $[i\gamma^{\mu}\partial_{\mu}-m]_{\lambda\beta}(-i)[\theta(x^0)\psi_{\beta}(x)\bar{\psi}_{\alpha}(0)-\theta(-x^0)\bar{\psi}_{\alpha}(0)\psi_{\beta}(x)]$ into
$[i\gamma^{\mu}\partial_{\mu}-m]_{\lambda\beta}(-i)[\theta(x^+)\psi_{\beta}(x)\bar{\psi}_{\alpha}(0)-\theta(-x^+)\bar{\psi}_{\alpha}(0)\psi_{\beta}(x)]$. Thus with  (\ref{7.3a}) and (\ref{7.6a}) both being derived from these q-number Dirac operator derivative relations,  they are completely equivalent. We thus establish  that for time-ordered operator products light-front quantization is instant-time quantization, with their apparent differences only occurring because of the restriction to equal times caused by the action of the time derivatives in the Dirac operator derivatives. In the derivation of (\ref{7.6a}) we note that in the instant-time time-ordered product the fermions are four-component spinors. Thus to match them on the light-front side we would also need four-component spinors. We would thus need both good and bad fermions, and thus see the specific role played by the bad fermion sector in establishing the equivalence of instant-time and light-front quantization for time-ordered products of fermion operators. As we thus see, the structure given in (\ref{7.2a}) follows from the structure of both the good and bad fermion anticommutation relations.

In order to extract out any possible observational consequences of the singular $\tfrac{i}{4}\gamma^+_{\beta\alpha}\delta(x^+)\epsilon(x^-)\delta(x^1)\delta(x^2)$ term that is present in  (\ref{7.2a}) we note that this term only takes support at $x^+=0$, $x^1=0$, $x^2=0$, and thus only at $x^+x^--(x^1)^2-(x^2)^2=0$, i.e., only on light-front light cone.  While $x^-$ is not fixed if $x^+=0$, $x^1=0$, $x^2=0$, nonetheless it has no effect on $x^+x^--(x^1)^2-(x^2)^2$ and thus we can take it to be zero too. We shall thus refer to the singular $\tfrac{i}{4}\gamma^+_{\beta\alpha}\delta(x^+)\epsilon(x^-)\delta(x^1)\delta(x^2)$ term as a tip of the light cone singularity. Thus for any process in which $x^+$ is not equal to zero (such as, for instance,  timelike scattering from one spacetime point to another, processes where $x^+> 0$) this tip of the light cone singularity makes no contribution. 

\begin{figure}[H]
\begin{center}
\includegraphics[scale=0.13]{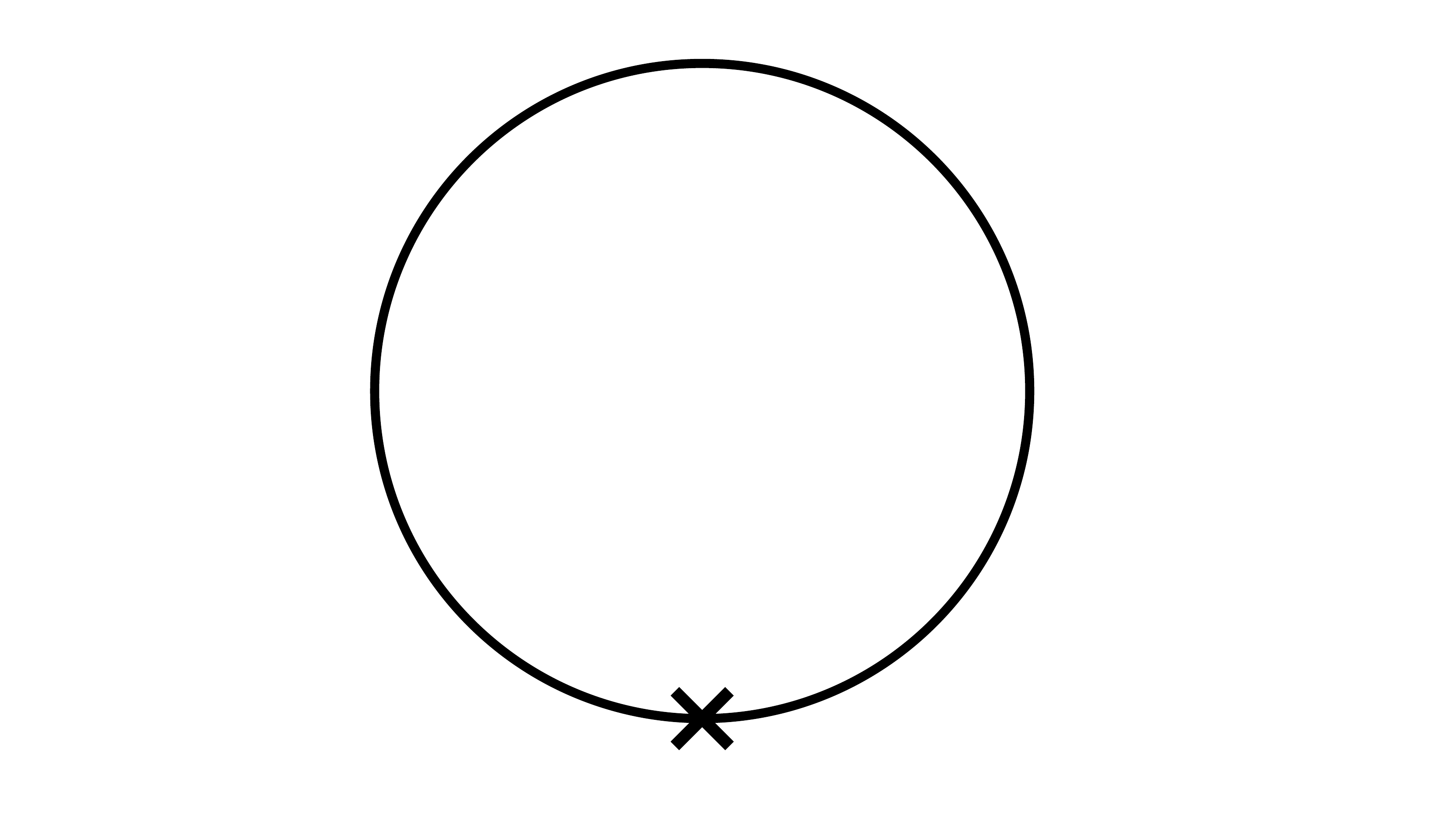}
\caption{ $\langle\Omega|\bar{\psi}(0)\psi(0)|\Omega\rangle$ tadpole graph}
\label{undressedtadpole}
\end{center}
\end{figure}

The one place where this singular term could contribute is if $x^{\mu}=0$, i.e., in light-front vacuum graphs such as $-i\langle \Omega|\psi_{\beta}(0)\bar{\psi}_{\alpha}(0)|\Omega\rangle$, a so-called vacuum tadpole graph of the type exhibited in Fig. \ref{undressedtadpole} that would arise in a Yukawa-coupled $\lambda \phi\bar{\psi}\psi$ theory, a graph  in which a scalar field brings zero four-momentum into the fermion loop at the point designated by the cross. As noted in  \cite{Mannheim2019a,Mannheim2019b}, we can construct such vacuum graphs by taking the $x^{\mu}\rightarrow 0$ limit of $-i\langle \Omega|T[\psi_{\beta}(x)\bar{\psi}_{\alpha}(0)]|\Omega\rangle$,  i.e., we can use the spacetime coordinate $x^{\mu}$ as a regulator. On taking the limit we obtain 
\begin{align}
&-i\langle \Omega|T[\psi_{\beta}(x)\bar{\psi}_{\alpha}(0)]|\Omega\rangle \rightarrow
i\langle \Omega|\bar{\psi}_{\alpha}(0)\psi_{\beta}(0)]|\Omega\rangle
-i\theta(0^+)\langle \Omega|[\psi_{\beta}(0)\bar{\psi}_{\alpha}(0)+\bar{\psi}_{\alpha}(0)\psi_{\beta}(0)]\Omega\rangle
\nonumber\\
&=i\langle \Omega |\bar{\psi}_{\alpha}(0)\psi_{\beta}(0)|\Omega\rangle
-i\theta(0^+)\gamma^0_{\nu\alpha}\langle \Omega |[\psi_{\beta}(0)(\Lambda^++\Lambda^-)\psi^{\dagger}_{\nu}(0)+\psi^{\dagger}_{\nu}(0)(\Lambda^++\Lambda^-)\psi_{\beta}(0)]|\Omega\rangle
\nonumber\\
&=i\langle \Omega |\bar{\psi}_{\alpha}(0)\psi_{\beta}(0)|\Omega\rangle
-i\theta(0^+)\gamma^0_{\nu\alpha}\langle \Omega |[\Big{\{}\psi^{(+)}_{\beta}(0),[\psi_{(+)}^{\dagger}]_{\nu}(0)\Big{\}}+\Big{\{}\psi^{(-)}_{\beta}(0),[\psi_{(-)}^{\dagger}]_{\nu}(0)\Big{\}}]|\Omega\rangle.
\label{7.7a}
\end{align}

Now the quantity $\psi_{\beta}(0)\bar{\psi}_{\alpha}(0)$ has 16 components. We can thus develop it in terms of irreducible representations of the Lorentz group as a scalar, a pseudoscalar, a vector, an axial vector, and a rank two antisymmetric tensor (i.e., as  $I$, $\gamma^5$, $\gamma^{\mu}$, $\gamma^{\mu}\gamma^5$ and $\gamma^{\mu}\gamma^{\nu}-\gamma^{\nu}\gamma^{\mu}$). However if Lorentz invariance is not to be broken in the vacuum, we can only allow the scalar and pseudoscalar.  Both involve taking a trace over the spinor indices, and since the discussion is equivalent for both we shall restrict to the scalar. Now according to  (\ref{1.14a}) and (\ref{1.19a}) the anticommutators that appear on the last line of (\ref{7.7a}) are proportional to $\Lambda^+$ and $\Lambda^-$. However both  
$\gamma^0\Lambda^+$ and $\gamma^0\Lambda^-$ are traceless. Thus in taking the trace of the last line in (\ref{7.7a}) the $\theta(0^+)$ dependent term drops out identically. Similarly, with $\gamma^+$ also being traceless, the $\gamma^+$ dependent term drops out of the trace of (\ref{7.2a}). Thus all that is left from the traces of (\ref{7.2a}) and (\ref{7.7a}) is 
\begin{align}
&\eta^{\alpha\beta}i\langle \Omega |\bar{\psi}_{\alpha}(0)\psi_{\beta}(0)|\Omega\rangle=i\langle \Omega |\bar{\psi}(0)\psi(0)|\Omega\rangle
=\frac{2}{(2\pi)^4}\int dp_+dp_1dp_2dp_-\frac{4m}{4p_+p_--(p_1)^2-(p_2)^2-m^2+i\epsilon}.
\label{7.8a}
\end{align}
Consequently, the tip of the light cone singularity drops out of the trace, leaving us with the one-loop Feynman diagram tadpole graph. 

By coordinate invariance this is exactly the same expression as the one associated with the fermion one-loop tadpole graph in instant-time quantization, and their equivalence is thus established. However, as noted in \cite{Mannheim2019a,Mannheim2019b} the way that the equivalence is actually established is due to a circle at infinity contribution in the complex light-front  $p_+$ plane (only one power of $p_+$ in the denominator of (\ref{7.8a})), a contribution that has no counterpart in the instant-time case as the circle in the complex instant-time  $p_0$ plane is suppressed (two powers of $p_0$ in the $(p_0)^2-(p_1)^2-(p_2)^2-(p_3)^2$ term in the denominator). The equivalence is thus prior to doing a contour integral with only the full pole plus circle contribution on an instant-time contour mapping into the full pole plus circle contribution on a light-front contour, with there being no separate mapping of pole into pole or circle into circle. Since (\ref{7.8a}) only differs from the analogous scalar field case though the factor of $4m$, discussion of the fermion loop tadpole graph is identical to discussion of the $\lambda \phi^3$ scalar field tadpole loop described in \cite{Mannheim2019a,Mannheim2019b}, where it was shown that the light-front and instant-time scalar field tadpoles, and thus now the light-front and instant-time fermion field tadpoles,  are indeed identical.

With the tip of the light cone singularity (where $x^+=0$) not contributing to the vacuum graph, and with it not contributing to non-vacuum graphs (where $x^+\neq 0$), we see that despite its presence,  the tip of the light cone singularity is not observable. Consequently, despite the presence of tip of the light cone singularities, in the fermionic sector light-front  quantized diagrams and instant-time quantized diagrams in both the non-vacuum and vacuum cases are equal. Since this result is derived by using light-front fermion anticommutation relations, and since these anticommutation relations themselves can be derived from instant-time anticommutation relations, we again see that light-front quantization is instant-time quantization. 

\section{Extension to Interacting Theories}
\label{S8}
\subsection{The Lehmann Representation For Commutators}
\label{S8a}

Having now seen the equivalence of instant-time quantization and light-front quantization in the free field theory case we comment briefly on how these results generalize to the interacting case both for commutators and Feynman diagrams.  For commutators (and analogously for anticommutators) the generalization is given by the Lehmann representation, an exact all-order relation in quantum field theory.  The Lehmann representation  is derived using only some very basic requirements that are thought to occur in any quantum field theory, requirements that do not get modified by the renormalization procedure. These requirements are Poincare invariance and the existence of Hermitian momentum generators $P_{\mu}$, generators that because of their  Hermiticity possess eigenstates that are complete and eigenvalues that are all real. The Hamiltonian is taken to possess a state of lowest energy, a  vacuum state that is taken to be unique. As such, these requirements make no reference to any dynamical equation that might be obeyed by the quantum fields of interest, while also not being restricted to perturbation theory. Moreover, for the purposes of actually deriving the Lehmann representation there is no need to actually determine any of the eigenvalues of the momentum generators or construct any of the associated eigenstates. All that matters is that they exist, and that must be the case if the momentum generators are Hermitian.

For the instant-time case first, following the discussion of the Lehmann representation given in \cite{Bjorken1965} we introduce the momentum generators $P_{\mu}=\int dx^1dx^2dx^3 T^{0}_{\phantom{0}\mu}$, with a Hermitian scalar field then transforming according to 
\begin{eqnarray}
[P_{\mu},\phi(x)]=-i\partial_{\mu}\phi(x),\quad \phi(x)=e^{iP\cdot x}\phi(0)e^{-iP\cdot x}.
\label{8.1}
\end{eqnarray}
We introduce a complete set of eigenstates $|p^n_{\mu}\rangle$ of the $P_{\mu}$ momentum generators, and other than the vacuum $|\Omega\rangle$ (an eigenstate of $P_{\mu}$ with all $p^n_{\mu}=0$), all the other states lie above the vacuum and have eigenvalues $p^n_{\mu}$ with  positive $p_0^n$ and non-negative $p^n_{\mu}p^{\mu}_n$, with the matrix element of $\phi(x)$ between $\langle \Omega|$ and  $|p_{\mu}^n\rangle$ being of the form
\begin{eqnarray}
&&\langle \Omega|\phi(x)|p^n_{\mu}\rangle=\langle \Omega|\phi(0)|p^n_{\mu}\rangle e^{-ip_{\mu}^n\cdot x},\quad \langle p_{\mu}^n|\phi(x)|\Omega\rangle=\langle p_{\mu}^n|\phi(0)|\Omega\rangle e^{ip_{\mu}^n\cdot x}.
\label{8.2}
\end{eqnarray}
With the states obeying the closure relation 
\begin{eqnarray}
\sum_n|n\rangle\langle n|=I,
\label{8.3}
\end{eqnarray}
we can then write the two-point function as
\begin{eqnarray}
\langle \Omega |\phi(x)\phi(y)|\Omega\rangle=\sum_n|\langle \Omega |\phi(0)|p_{\mu}^n\rangle |^2e^{-ip^n\cdot(x-y)}.
\label{8.4}
\end{eqnarray}
While the closure sum on $n$ given in (\ref{8.3}) must include the vacuum (in order to enforce $\sum_n|n\rangle\langle n|\Omega\rangle=|\Omega\rangle$), the vacuum does not contribute in the sum on $n$ in (\ref{8.4}), since the uniqueness of the vacuum requires that $\langle \Omega |\phi(0)|\Omega\rangle$ be zero as one otherwise would have  spontaneous symmetry breaking and a degenerate vacuum.

We now  introduce the instant-time ($IT$) spectral function
\begin{eqnarray}
\rho(q_{\mu},IT)=(2\pi)^3\sum_n\delta^4(p_{\mu}^n-q_{\mu})|\langle \Omega |\phi(0)|p_{\mu}^n\rangle |^2=\rho(q^2,IT)\theta(q_0).
\label{8.5}
\end{eqnarray}
Because of Lorentz invariance the spectral function is only a function of $q^2$, and with all the $p_0^n$ being positive, $\rho(q_{\mu},IT)$ can be written as $\rho(q_{\mu},IT)=\rho(q^2,IT)\theta(q_0)$, with $\rho(q^2,IT)$  vanishing for $q^2<0$ since all the $p^n_{\mu}$ have non-negative $p_{\mu}^np^{\mu}_n$. In terms of $\rho(q_{\mu},IT)$ we obtain the two-point function
Lehmann representation 
\begin{eqnarray}
\langle \Omega |\phi(x)\phi(y)|\Omega\rangle&=&\frac{1}{(2\pi)^3}\int d^4q \rho(q^2,IT)\theta(q_0)e^{-iq\cdot(x-y)}
\nonumber\\
&=&\frac{1}{(2\pi)^3}\int_0^{\infty}d\sigma^2\rho(\sigma^2,IT)\int d^4q \theta(q_0)\delta(q^2-\sigma^2)e^{-iq\cdot(x-y)},
\label{8.6}
\end{eqnarray}
where $\sigma$ is a mass parameter. Thus with $\epsilon(q_0)=\theta(q_0)-\theta(-q_0)$, the vacuum matrix element of the commutator is given by 
\begin{eqnarray}
\langle \Omega |[\phi(x),\phi(y)]|\Omega\rangle=\frac{1}{(2\pi)^3}\int_0^{\infty}d\sigma^2\rho(\sigma^2,IT)\int d^4q \epsilon(q_0)\delta(q^2-\sigma^2)e^{-iq\cdot(x-y)}.
\label{8.7}
\end{eqnarray}

Now in (\ref{2.2a}) we expressed the free theory $i\Delta(IT;x)$ for a scalar field of mass $m$ as an on-shell three-dimensional integral. On introducing a delta function we can rewrite it as a still on-shell four-dimensional integral. And relabeling it as $i\Delta(IT,FREE;x,m^2)$ we rewrite (\ref{2.2a}) as
\begin{eqnarray}
i\Delta(IT,FREE;x-y,m^2)=\frac{1}{(2\pi)^3}\int d^4q\epsilon(q_0)\delta(q^2-m^2)e^{-iq\cdot(x-y)},
\label{8.8}
\end{eqnarray}
with its evaluation for $(x-y)^2>0$, $(x-y)^2=0$ and $(x-y)^2<0$ being given in (\ref{2.5a}).
Labeling the full commutator term on the left-hand side of (\ref{8.7}) as $i\Delta(IT,FULL;x-y)$, we can thus rewrite (\ref{8.7}) as
\begin{eqnarray}
i\Delta(IT,FULL;x-y)=\int_0^{\infty} d\sigma^2 \rho(\sigma^2,IT)i\Delta(IT,FREE;x-y,\sigma^2),
\label{8.9}
\end{eqnarray}
with the range of $\sigma^2$ being restricted to $(0,\infty)$ since $\rho(\sigma^2,IT)$ vanishes outside the light cone.
With recognize (\ref{8.9}) as  the Lehmann representation for the $x^0\neq y^0$ commutator  \cite{Bjorken1965}.

For the light-front ($LF$) case we again assume Poincare invariance and Hermiticity of the momentum generators. With the requirement that all $p_{\mu}^np_n^{\mu}$ be non-negative, it follows that the instant-time momenta obey $(p^n_0)^2-(p^n_3)^2-(p^n_1)^2-(p^n_2)^2\geq 0$, with light-front momenta thus obeying $4p^n_+p^n_--(p^n_1)^2-(p^n_2)^2\geq 0$. Then with $p_0^n>|p_3^n|$ it follows that $p_+^n=p_0^n+p_3^n$ and $p_-^n=p_0^n-p_3^n$ are both positive. Boundedness from below of the instant-time Hamiltonian thus entails boundedness from below of the light-front Hamiltonian  as well. 

To now establish the Lehmann representation in the light-front case we need to write the three-dimensional free light-front theory $x^+\neq y^+$ commutator $ i\Delta(LF,FREE;x-y,m^2)$ given in (\ref{2.4a}) in a four-dimensional form. We anticipate that it will be the analog of (\ref{8.8}) and check to see if this is the case. We thus set
\begin{eqnarray}
&&i\Delta(LF,FREE;x-y,m^2)=\frac{2}{(2\pi)^3}\int_{-\infty}^{\infty} dq_1dq_2dq_+dq_-\epsilon(q_+)\delta(4q_+q_--F_q^2)
\nonumber\\
&&\times e^{-iq_+(x^+-y^+)-iq_-(x^--y^-)-iq_1(x^1-y^1)-iq_2(x^2-y^2)},
\label{8.10}
\end{eqnarray}
where $F_q^2=(q_1)^2+(q_2)^2+m^2$, and where the factor of  $(-g)^{-1/2}=2$ was introduced in Sec. \ref{S1}. With the delta function term requiring that $q_+q_-$ be positive, then given the $\epsilon(q_+)$ term  the integration breaks up into two pieces:
\begin{eqnarray}
i\Delta(LF,FREE;x-y,m^2)&=&\frac{2}{(2\pi)^3}\int_{-\infty}^{\infty} dq_1dq_2\int_0^{\infty} dq_+dq_-\delta(4q_+q_--F_q^2)e^{-iq\cdot(x-y)}
\nonumber\\
&-&\frac{2}{(2\pi)^3}\int_{-\infty}^{\infty} dq_1dq_2\int _{-\infty}^0dq_+dq_-\delta(4q_+q_--F_q^2)e^{-iq\cdot(x-y)}.
\label{8.11}
\end{eqnarray}
Setting $q_{\mu}=-q_{\mu}$ in the second integral then yields
\begin{eqnarray}
i\Delta(LF,FREE;x-y,m^2)&=&\frac{2}{(2\pi)^3}\int_{-\infty}^{\infty} dq_1dq_2\int_0^{\infty} dq_+dq_-\delta(4q_+q_--F_q^2)e^{-iq\cdot(x-y)}
\nonumber\\
&-&\frac{2}{(2\pi)^3}\int_{-\infty}^{\infty} dq_1dq_2\int _0^{\infty}dq_+dq_-\delta(4q_+q_--F_q^2)e^{iq\cdot(x-y)}.
\label{8.12}
\end{eqnarray}
Finally, doing the $q_+$ integration yields
\begin{eqnarray}
i\Delta(LF,FREE;x-y,m^2)&=&\frac{2}{(2\pi)^3}\int_{-\infty}^{\infty}dq_1\int_{-\infty}^{\infty}dq_2\int_{0}^{\infty}\frac{dq_-}{4q_-}
\nonumber\\
&&\times
\left[e^{-i(F_q^2x^+/4q_-+q_-x^-+q_1x^1+q_2x^2)}-e^{i(F_q^2x^+/4q_-+q_-x^-+q_1x^1+q_2x^2)}\right].
\label{8.13}
\end{eqnarray}
We recognize (\ref{8.13}) as (\ref{2.4a}), and thus confirm the validity of (\ref{8.10}), with its evaluation for $(x-y)^2>0$ being given in (\ref{2.9a}), for $(x-y)^2=0$ being given in (\ref{2.10a}), and with it vanishing for $(x-y)^2<0$. 

Defining now a light-front spectral function
\begin{eqnarray}
\rho(q_{\mu},LF)=\frac{(2\pi)^3}{2}\sum_n\delta^4(p_{\mu}^n-q_{\mu})|\langle \Omega |\phi(0)|p_{\mu}^n\rangle |^2=\rho(q^2,LF)\theta(q_+),
\label{8.14}
\end{eqnarray}
then with (\ref{8.4}) also holding in light-front coordinates, the $x^+\neq y^+$ light-front coordinate commutator is given by
\begin{eqnarray}
\langle \Omega |[\phi(x),\phi(y)]|\Omega\rangle=\frac{2}{(2\pi)^3}\int_0^{\infty}d\sigma^2\rho(\sigma^2,LF)\int d^4q \epsilon(q_+)\delta(q^2-\sigma^2)e^{-iq\cdot(x-y)}.
\label{8.15}
\end{eqnarray}
Thus in the light-front case the Lehmann representation takes the form 
\begin{eqnarray}
i\Delta(LF,FULL;x-y)=\int_0^{\infty} d\sigma^2 \rho(\sigma^2,LF)i\Delta(LF,FREE;x-y,\sigma^2).
\label{8.16}
\end{eqnarray}

We recognize the light-front (\ref{8.16}) as being completely analogous to the instant-time Lehmann representation given in (\ref{8.9}), and thus anticipate that one can be transformed into the other. We will need to transform both the free $i\Delta(IT,FREE;x-y,\sigma^2)$ and the spectral function. For the free $i\Delta(IT,FREE;x-y,\sigma^2)$ given in (\ref{8.8}) we  set $p_0=p_++p_-$, $p_3=p_+-p_-$. This transforms $(p_0)^2-(p_3)^2-(p_1)^2-(p_2)^2-m^2$ into $4p_+p_--(p_1)^2-(p_2)^2-m^2$. From the delta function constraint it follows that $p_+p_-$ is positive, with $p_+$ and $p_-$ thus having  the same sign. Consequently, with  $\epsilon(p_0)$ transforming into $\epsilon(p_++p_-)$ it follows that $\epsilon(p_++p_-)\delta(p^2-m^2)$ is equal to $\epsilon(p_+)\delta(p^2-m^2)$ alone. Then, after making an analogous transformation on $x^{\mu}-y^{\mu}$ of the form $x^0-y^0=(x^+-y^++x^--y^-)/2$, $x^3-y^3=(x^+-y^+-x^-+y^-)/2$, the equivalence of $i\Delta(IT,FREE;x-y,\sigma^2)$ as given in (\ref{8.8}) and $i\Delta(LF,FREE;x-y,\sigma^2)$ as given in (\ref{8.10}) is established. 

To compare the spectral functions we first write the free theory Fock expansions of (\ref{2.1a}) and (\ref{2.3a}) in a four-dimensional form. We introduce $A(IT;\vec{p}=(p_1,p_2,p_3))=(2E_p)^{1/2}a(IT;\vec{p})$ in the instant-time case \cite{Bjorken1965} and $A(LF;\vec{p}=(p_1,p_2,p_-))=(4p_-)^{1/2}a(LF;\vec{p})$ in the light-front case, both as confined to their respective mass shells, $(p_0)^2=E_p^2$, $4p_+p_-=F_p^2$. These creation and annihilation operators obey
\begin{eqnarray}
&&[A(IT;\vec{p}),A^{\dagger}(IT;\vec{p}^{\prime})]=2E_p\delta(p_1-p_1^{\prime})\delta(p_2-p_2^{\prime})\delta(p_3-p_3^{\prime}),
\nonumber\\
&&[A(LF;\vec{p}),A^{\dagger}(LF;\vec{p}^{\prime})]=4p_-\tfrac{1}{2}\delta(p_1-p_1^{\prime})\delta(p_2-p_2^{\prime})\delta(p_--p_-^{\prime}).
\label{8.17}
\end{eqnarray}
The Fock space expansions thus take the form 
\begin{eqnarray}
\phi(IT;x^0,x^1,x^2,x^3)&=&\frac{1}{(2\pi)^{3/2}}\int d^4p \delta((p_0)^2-E_p^2)\theta(p_0)\left[A(IT;\vec{p})e^{-ip\cdot x}
+A^{\dagger}(IT;\vec{p})e^{+ip\cdot x}\right]
\nonumber\\
&=&\frac{1}{(2\pi)^{3/2}}\int d^4p \delta((p_0)^2-E_p^2)\left[\theta(p_0)A(IT;\vec{p})
+\theta(-p_0)A^{\dagger}(IT;-\vec{p})\right]e^{-ip\cdot x}.
\label{8.18}
\end{eqnarray}
\begin{eqnarray}
\phi(LF;x^+,x^1,x^2,x^-)&=&\frac{2}{(2\pi)^{3/2}}\int_{-\infty}^{\infty}dp_1dp_2\int_0^{\infty}dp_+dp_- \delta(4p_+p_--F_p^2)A(LF;\vec{p})e^{-ip\cdot x}
\nonumber\\
&+&\frac{2}{(2\pi)^{3/2}}\int_{-\infty}^{\infty}dp_1dp_2\int_0^{\infty}dp_+dp_- \delta(4p_+p_--F_p^2)A^{\dagger}(LF;\vec{p})e^{ip\cdot x}
\nonumber\\
&=&\frac{2}{(2\pi)^{3/2}}\int_{-\infty}^{\infty}dp_1dp_2\int_0^{\infty}dp_+dp_- \delta(4p_+p_--F_p^2)A(LF;\vec{p})e^{-ip\cdot x}
\nonumber\\
&+&\frac{2}{(2\pi)^{3/2}}\int_{-\infty}^{\infty}dp_1dp_2\int_{-\infty}^0 dp_+dp_- \delta(4p_+p_--F_p^2)A^{\dagger}(LF;-\vec{p})e^{-ip\cdot x}
\nonumber\\
&=&\frac{2}{(2\pi)^{3/2}}\int d^4p \delta(4p_+p_--F_p^2)\left[\theta(p_+)A(LF;\vec{p})
+\theta(-p_+)A^{\dagger}(LF;-\vec{p})\right]e^{-ip\cdot x}.
\label{8.19}
\end{eqnarray}
The utility of these expressions is that they show that $A(IT;\vec{p})$ and $A(LF;\vec{p})$ are Lorentz scalars. ($\int d^3p/2E_p\times 2E_p\delta^3(\vec{p}-\vec{p}^{\prime})=1$, $2\int dp_1dp_2\int_0^{\infty}dp_-/4p_-\times 4p_-\tfrac{1}{2}\delta(p_1-p_1^{\prime})\delta(p_2-p_2^{\prime})\delta(p_--p_-^{\prime})=1$.)

As noted in \cite{Mannheim2019a}, with the transformation $x^0\rightarrow x^0+x^3$, $x^3\rightarrow x^0-x^3$ being a general coordinate translation on the coordinates (and incidentally not a Lorentz transformation),  assuming the theory to be general coordinate invariant  we can transform $\phi(IT;x)$ to $\phi(LF;x)$ at the operator level by introducing the unitary translation operator 
\begin{eqnarray}
U(P_0,P_3)=\exp(ix^3P_0)\exp(ix^0P_3),
\label{8.20}
\end{eqnarray}
where the $P_{\mu}$ are momentum generators that effect $[P_{\mu},\phi(x)]=-i\partial_{\mu}\phi$. In order to apply this transformation to (\ref{8.18}),  we note that because we use $x^3\rightarrow x^0-x^3=x^-$ rather than $x^3\rightarrow x^3-x^0=-x^-$, then rather than reformulate everything in terms of   $x^3\rightarrow x^3-x^0$, we shall restrict to theories in which the action is invariant under $x^3\rightarrow -x^3$ (i.e., actions with only even powers of $\phi$), so that we can set  $\phi(IT;x^0,x^1,x^2,x^3)=\phi(IT;x^0,x^1,x^2,-x^3)$. On now applying (\ref{8.20})  to the left-hand side of (\ref{8.18}) we obtain
\begin{align}
U\phi(IT;x^0,x^1,x^2,x^3)U^{-1}=U\phi(IT;x^0,x^1,x^2,-x^3)U^{-1}=\phi(IT;x^0+x^3,x^1,x^2,x^0-x^3)=\phi(LF;x^+,x^1,x^2,x^-).
\label{8.21}
\end{align}
On substituting $p_0=p_++p_-$, $p_3=p_+-p_-$ into the right-hand side of (\ref{8.18}) we obtain 
\begin{eqnarray}
\phi(LF;x^+,x^1,x^2,x^-)&=&\frac{2}{(2\pi)^{3/2}}\int d^4p \delta(4p_+p_--F_p^2)
\nonumber\\
&&\times\left[\theta(p_++p_-)UA(IT;\vec{p})U^{-1}
+\theta(-p_+-p_-)UA^{\dagger}(IT;-\vec{p})U^{-1}\right]e^{-ip\cdot x},
\label{8.22}
\end{eqnarray}
as now written in light-front coordinates. Then, since $\delta(4p_+p_--F_p^2)\theta(\pm (p_++ p_-))=\delta(4p_+p_--F_p^2)\theta(\pm p_+)$, we obtain
\begin{align}
\phi(LF;x^+,x^1,x^2,x^-)=\frac{2}{(2\pi)^{3/2}}\int d^4p \delta(4p_+p_--F_p^2)\left[\theta(p_+)UA(IT;\vec{p})U^{-1}
+\theta(-p_+)UA^{\dagger}(IT;-\vec{p})U^{-1}\right]e^{-ip\cdot x}.
\label{8.23}
\end{align}
Finally, comparing with (\ref{8.19}) and recalling that $U$ is unitary, we obtain
\begin{eqnarray}
UA(IT; \vec{p})U^{-1}=A(LF;\vec{p}),\quad UA^{\dagger}(IT; -\vec{p})U^{-1}=A^{\dagger}(LF;-\vec{p}),\quad UA^{\dagger}(IT; \vec{p})U^{-1}=A^{\dagger}(LF;\vec{p}).
\label{8.24}
\end{eqnarray}
The instant-time and light-front scalar field and the associated creation and annihilation operators are thus unitarily equivalent. With 
\begin{align}
A(IT; \vec{p})|\Omega(IT)\rangle=0,&\quad A^{\dagger}(IT;\vec{p})|\Omega(IT)\rangle=|p_{\mu}(IT)\rangle,
\nonumber\\
A(LF; \vec{p})|\Omega(LF)\rangle=0,&\quad A^{\dagger}(LF;\vec{p})|\Omega(LF)\rangle=|p_{\mu}(LF)\rangle,
\label{8.25}
\end{align}
we obtain
\begin{eqnarray}
&&U|\Omega(IT)\rangle=|\Omega(LF)\rangle,\quad U|p_{\mu}(IT)\rangle=|p_{\mu}(LF)\rangle,\quad
U\phi(IT;0)U^{-1}=\phi(LF;0).
\label{8.26}
\end{eqnarray}
Given (\ref{8.26}) we thus establish the equivalence of the instant-time and light-front spectral functions given in (\ref{8.5}) and (\ref{8.14}) and thus of the Lehmann representations in the free theory case. Since translations are general coordinate transformations, our result follows from the general coordinate invariance of the scalar field theory. And as is standard in quantum theory, invariance under translations entails invariance under the unitary transformations associated with the translation generators, with instant-time quantization and light--front quantization thus being unitarily equivalent. 

Finally, for the full interacting Lehmann representation, we can transform with the unitary $U(P_0,P_3)$ operator to all orders since the full fields and full momentum operators obey the relation $[P_{\mu},\phi]=-i\partial_{\mu}\phi$ given in (\ref{8.1}) that was initially  used for establishing the Lehmann representation in the first place. The general coordinate invariance of the interacting theory, which we take to be the case,  then establishes the unitary equivalence of the interacting instant-time and light-front theories, with (\ref{8.26}) actually holding to all orders and not just for the free theory. Consequently, we see that 
\begin{align}
\langle \Omega(IT)|\phi(IT;x^0,x^1,x^2,x^3))|p^n_{\mu}(IT)\rangle&=\langle \Omega(IT)|U^{\dagger}U\phi(IT;x^0,x^1,x^2,x^3))U^{\dagger}U|p^n_{\mu}(IT)\rangle
\nonumber\\
&=\langle \Omega(LF)|\phi(LF;x^+,x^1,x^2,x^-))|p^n_{\mu}(LF)\rangle.
\label{8.27}
\end{align}
We thus establish the equivalence of the interacting spectral functions, and thus the equivalence of the full Lehmann representations. We thus see that to all orders the vacuum expectation values of the commutators in the instant-time and light-front cases are the same. Thus by working with the commutators at unequal $x^0-y^0$ and unequal $x^+-y^+$ we can establish the equivalence of instant-time and light-front quantization to all orders. This generalizes the free scalar field theory result that had been given in Sec. \ref{S2} and again shows the centrality of the general unequal time commutators. (An analogous Lehmann representation also exists for anticommutators but we do not discuss it here.)

\subsection{Feynman Diagrams}
\label{S8b}

For Feynman diagrams we note that the analysis of Sec. \ref{S7} shows the equivalence of instant-time and light-front Feynman propagators for a free fermion theory. An analogous analysis given in \cite{Mannheim2019a} establishes the same equivalence for free bosons. However, once we have the equivalence for free theory propagators, via the Dyson-Wick expansion we then have the same equivalence in the interacting case order by order in perturbation theory. The free theory analysis thus generalizes to the interacting case. To buttress this result we note that in \cite{Mannheim2019a} we reached the same conclusion via an analysis of the path integral representation of propagators. Specifically, we note that as well as being writable as integrals over momentum variables in momentum space, Feynman diagrams can also be written as path integrals over classical field paths in coordinate space. Both of these types of  integrals are integrals over purely classical variables with no reference to quantum operators. (One does of course need quantum operators in order to identify the propagator as a matrix element of the quantum operators in the first place, but once one has the matrix element one only has to deal with an integration over classical momentum variables.) Since both Feynman diagrams and path integrals only involve integrations over classical variables, one can transform both of them from the instant-time form to the light-front form just by a change of variables, either momentum variables ($p_0=p_++p_-$, $p_3=p_+-p_-$) or coordinate variables ($x^0-y^0=(x^+-y^++x^--y^-)/2$, $x^3-y^3=(x^+-y^+-x^-+y^-)/2$), with the equivalence then directly following.

Establishing this equivalence for Feynman propagators and for Feynman path integrals is actually a lot more straightforward than establishing the equivalence for the Lehmann representation as one does not have to deal with quantum operators but only with classical variables. Nonetheless, with the Lehmann representation holding for the two-point function and not just for the commutator, the Lehmann representation also holds for time-ordered products, relating the full interacting  $D(x^{\mu}-y^{\mu})=-i\langle \Omega |T[\phi(x)\phi(y)]|\Omega\rangle$ propagator to the free propagator according to
\begin{align}
D(IT,FULL;x-y)&=\int_0^{\infty} d\sigma^2 \rho(\sigma^2,IT)D(IT,FREE;x-y,\sigma^2),
\nonumber\\
D(LF,FULL;x-y)&=\int_0^{\infty} d\sigma^2 \rho(\sigma^2,LF)D(LF,FREE;x-y,\sigma^2),
\label{8.28}
\end{align}
with the same instant-time and light-front spectral functions as given in (\ref{8.5}) and (\ref{8.14}), and with
\begin{align}
&D(IT,FREE;x,\sigma^2)=\frac{1}{(2\pi)^4}\int dp_0dp_1dp_2dp_3 \frac{e^{-i(p_0x^0+p_1x^1+p_2x^2+p_3x^3)}}{(p_0)^2-(p_1)^2-(p_2)^2-(p_3)^2-\sigma^2+i\epsilon},
\nonumber\\
&D(LF,FREE;x,\sigma^2)=\frac{2}{(2\pi)^4}\int dp_+dp_1dp_2dp_- \frac{e^{-i(p_+x^++p_1x^1+p_2x^2+p_-x^-)}}{4p_+p_--(p_1)^2-(p_2)^2-\sigma^2+i\epsilon}.
\label{8.29}
\end{align}
With the free propagators transforming into each other under the substitutions $p_0=p_++p_-$, $p_3=p_+-p_-$, $x^0=(x^++x^-)/2$, $x^3=(x^+-x^-)/2$, and with the spectral functions transforming into each other, one can even establish the equivalence of all-order instant-time and light-front time-ordered products with needing to make any reference to perturbation theory at all. 

Since Feynman diagrams diverge we need to introduce a renormalization procedure. Because any given instant-time Feynman diagram and its light-front counterpart are equal they diverge at the same rate in the ultraviolet, and thus can be renormalized by the same set of counterterms. Thus as long as we use general coordinate invariant counterterms, which we of course do, the equivalence of instant-time and light-front Feynman diagrams continues  to hold after renormalization. In formulating the Dyson-Wick expansion we also have to deal with products of fields at the same spacetime points, ill-defined objects that we treat by normal ordering. In \cite{Mannheim2019a,Mannheim2019b} it was shown that one can construct the $\langle \Omega|\phi(0)\phi(0)|\Omega\rangle$ vacuum tadpole graph as the $x^{\mu}\rightarrow 0$ limit of the time-ordered product $\langle \Omega|T[\phi(x)\phi(0)]|\Omega\rangle$. Consequently,  the same reasoning that ensures that instant-time and light-front Feynman diagrams are equal also ensures that the short-distance behaviors of matrix elements of instant-time and light-front products of fields at the same spacetime points are also equal. Then, with the $U|\Omega(IT)\rangle=|\Omega(LF)\rangle$ relation given in (\ref{8.26})  holding to all orders, normal ordering instant-time products of fields with respect to $|\Omega(IT)\rangle$ is the same as normal ordering light-front products with respect to $|\Omega(LF)\rangle$. The normal-ordering prescriptions in the two cases are thus equivalent. 

While establishing the equivalence of instant-time and light-front Feynman diagrams is straightforward in principle, in practice when it comes to actually evaluating the diagrams there are difficulties on the light-front side associated with zero modes (modes with $p_-=0$), difficulties that have no instant-time counterpart. Handling these zero-mode difficulties requires some care, and some prescriptions for handling them have been developed in \cite{Mannheim2019a,Mannheim2019b}, with different prescriptions being needed for non-vacuum and vacuum diagrams. For both cases the prescriptions involve considering the coordinate dependence of objects such as the time-ordered product $\langle \Omega |T[\phi(x)\phi(0)]|\Omega\rangle$.  As long as $x^{\mu}$ is non-zero we use $x^{\mu}$ itself as a regulator and show that the $p_-\rightarrow 0$ limit is non-singular. However, this is not the case if $x^{\mu}=0$, the vacuum tadpole bubble case, and then we use the same regulator prescription  $\int d\alpha e^{i\alpha(A+i\epsilon)}=-1/i(A+i\epsilon)$ and $\int d\alpha e^{-i\alpha(A-i\epsilon)}=1/i(A-i\epsilon)$ as used in (\ref{2.7a}) above. The great advantage of this prescription is that it puts $p_-$ into an exponent, and then one can take the $p_-\rightarrow 0$ limit without difficulty. Thus by looking at Feynman diagrams in coordinate space rather than in momentum space we develop a procedure for handling zero modes that generalizes to the arbitrary interacting Feynman diagram.  Basically, to handle zero modes one must not start with an on-shell 3-dimensional formalism as then the $p_-\rightarrow 0$ limit is singular (cf. the on-shell Fock expansion given in (\ref{2.3a}) with its $1(4p_-)^{1/2}$ term). Rather, one must first go off-shell by introducing a Feynman contour so as to obtain a four-dimensional formalism. Then using Feynman's $i\epsilon$ prescription one can introduce the exponential regulators and evaluate  Feynman diagrams without any zero-mode problems. This prescription holds for any perturbative Feynman diagram to any order in interactions. As an explicit example, in \cite{Mannheim2019b} the first radiative correction to the vacuum tadpole bubble graph given in Fig. \ref{undressedtadpole} was examined in detail and no new zero-mode issues were found to arise.

In the study of \cite{Mannheim2019a} it was shown that for non-vacuum graphs one only has pole terms in both the  instant-time and light-front cases, with the pole term contributions coinciding in the two cases. Since there are only pole terms the on-shell light-front Hamiltonian formalism results are recovered. As shown in \cite{Mannheim2019a,Mannheim2019b}, and as  noted in Sec. \ref{S7} above, for the vacuum bubble there is also a circle at infinity contribution in the light-front case, even though there is none in the instant-time case. However, the net effect of poles in the instant-time vacuum bubble and poles and circle contributions in the light-front vacuum bubble case is that the net contributions in the two cases are the same. We had noted above that one can show the equivalence of instant-time and light-front Feynman diagrams by a straightforward change of variables. In such a change a contour integration in the complex $p_0$ plane transforms into a contour integration in the complex $p_+$ plane. The equivalence is thus established before one actually evaluates the various pole and circle at infinity contributions to the contour integral. The equivalence is thus is for the net contributions, i.e., poles and circles combined, and not for poles and circles separately. Since the pole contribution is an on-shell contribution, it is precisely in the pole contribution that one encounters the zero-mode problem (the on-shell light-front Fock space expansion given in (\ref{2.3a}) for instance possesses a zero-mode singularity). However, there is another way to evaluate the contour integral, one that avoids poles altogether, namely to use the $\alpha$ regulator prescription described above. Then one never has to consider pole terms and there is no zero-mode problem at all. 

That the light-front theory must be able to handle zero-mode problems can be understood from the Feynman path integral representation of Feynman diagrams. Such path integrals are integrals over paths in coordinate space, alone. Consequently, in them momentum space $p_-=0$ zero-mode issues are never encountered. However, any given Green's function has both a coordinate space path integral representation and a momentum space Feynman diagram  representation. Thus if formulated carefully as described above, one can avoid encountering zero-mode problems in momentum space Feynman diagrams. 

\section{Comparing the Instant-Time and Light-Front Hamiltonians}
\label{S9}

Unlike commutators, which are local objects defined at local coordinates $x^{\mu}$ and $y^{\mu}$,  Hamiltonian operators are global objects as they are integrals over all space of the energy-momentum tensor, with spatially asymptotic boundary conditions being needed in order to show that they are time independent. Now transformations such as $x^0\rightarrow x^0+x^3$,   $x^3\rightarrow x^0-x^3$ are spacetime-dependent translations, and are thus general coordinate transformations. To construct the energy-momentum tensor one varies a covariantized action with respect to the metric. Since the action is general coordinate invariant, the instant-time and light-front energy-momentum tensors constructed this way are coordinate equivalent, and in this local sense they are completely equivalent. As constructed, both the instant-time energy-momentum tensor and the light-front energy-momentum tensor obey $\partial_{\mu}T^{\mu\nu}=0$. The respective Hamiltonians are  then constructed as 
\begin{eqnarray}
H(IT)=\int dx^1dx^2dx^3T^0_{\phantom{0}0},\quad H(LF)=\tfrac{1}{2}\int dx^1dx^2dx^-T^+_{\phantom{+}+}.
\label{9.1}
\end{eqnarray}
Since $\partial_{\mu}T^{\mu}_{\phantom{\mu}\nu}=0$ the Hamiltonians obey
\begin{eqnarray}
&&\partial_0H(IT)=-\int dx^1dx^2dx^3\left[\partial_1T^{1}_{\phantom{1}0}+\partial_2T^{2}_{\phantom{2}0}+\partial_3T^{3}_{\phantom{3}0}\right],
\nonumber\\
&&\partial_+H(IT)=-\tfrac{1}{2}\int dx^1dx^2dx^-\left[\partial_1T^{1}_{\phantom{1}+}+\partial_2T^{2}_{\phantom{2}+}+\partial_-T^{-}_{\phantom{-}+}\right].
\label{9.2}
\end{eqnarray}
With both of the integrals on the right-hand sides of (\ref{9.2}) being asymptotic surface terms, to show that $H(IT)$ and $H(LF)$ are respectively independent of $x^0$ and $x^+$ requires different global boundary conditions (conditions that we assume to hold), namely asymptotic vanishing in $x^1$, $x^2$ and $x^3$ for $H(IT)$ and asymptotic vanishing in $x^1$, $x^2$ and $x^-$ for $H(LF)$. Now, as such, these  boundary conditions do not transform into each other because $x^1$, $x^2$ and $x^3$ can be reexpressed as $x^1$, $x^2$ and $(x^+-x^-)/2$ and not as $x^1$, $x^2$ and $x^-$. Moreover, the functions $T^{1}_{\phantom{1}0}$, $T^{2}_{\phantom{2}0}$ and  $T^{3}_{\phantom{3}0}$ are different from $T^{1}_{\phantom{1}+}$, $T^{2}_{\phantom{2}+}$ and  $T^{-}_{\phantom{-}+}$. (For fermions these light-front components even contain the non-local bad fermions \cite{Mannheim2019a}.) Thus even though one can transform the local energy-momentum tensors into each other one cannot transform the Hamiltonians into each other. In this respect then $H(IT)$ and $H(LF)$ are intrinsically different. 

As we noted in Sec. \ref{S1}, the canonical equal instant-time and equal light-front time commutators are also intrinsically different. However, as shown in \cite{Mannheim2019a}, it is this very difference that actually enables both instant-time and light-front momentum generators to obey $[P_{\mu},\phi]=-i\partial_{\mu}\phi$. Thus despite there being both global differences (asymptotic boundary conditions) and local differences (canonical commutators), nonetheless one does find some commonality, namely $H(IT)$ generates translations in $x^0$ and its light-cone $H(LF)$ counterpart generates translations in $x^+$, just as they should. Moreover, for the other momentum generators $P_3(IT)$ generates translations in $x^3$ and $P_-(LF)$ generates translations in $x^-$, while both $P_1(IT)$ and $P_1(LF)$ generate translations in $x^1$ and  both $P_2(IT)$ and $P_2(LF)$ generate translations in $x^2$.

To see just how equivalent $H(IT)$ and $H(LF)$ might be, we note that for a free scalar theory with $T_{\mu\nu}=\partial_{\mu}\phi\partial_{\nu}\phi-\tfrac{1}{2}g_{\mu\nu}(\partial^{\alpha}\phi\partial_{\alpha}\phi-m^2\phi^2)$, we can use the Fock expansions given in (\ref{2.1a}) and (\ref{2.3a}) to evaluate $H(IT)$ and $H(LF)$, and with the form for $T^{+}_{\phantom{+}+}$ given in \cite{Mannheim2019a} obtain
\begin{align}
H(IT)&=\int dx^1dx^2dx^3\tfrac{1}{2}\left[(\partial_0\phi)^2+\vec{\nabla}\phi\cdot \vec{\nabla} \phi+m^2\phi^2\right]
=\tfrac{1}{2}\int_{-\infty}^{\infty}dp_1dp_2dp_3\left[a^{\dagger}(\vec{p})a(\vec{p})+a(\vec{p})a^{\dagger}(\vec{p})\right]p_0,
\nonumber\\
H(LF)&=\tfrac{1}{2}\int dx^1dx^2dx^-\tfrac{1}{2}\left[(\partial_1\phi)^2+(\partial_2\phi)^2+m^2\phi^2\right]
=\tfrac{1}{2}\int_{-\infty}^{\infty} dp_1dp_2\int_{0}^{\infty}dp_-\left[a^{\dagger}(\vec{p})a(\vec{p})+a(\vec{p})a^{\dagger}(\vec{p})\right]p_+,
\label{9.3}
\end{align}
with the allowed values of $p_0=E_p$ and $p_+=F_p^2/4p_-$ being positive. In terms of the $A(IT;\vec{p})=(2E_p)^{1/2}a(IT;\vec{p})$ and $A(LF;\vec{p})=(4p_-)^{1/2}a(LF;\vec{p})$ operators introduced above we can write both the Hamiltonian and the other momentum generators in the covariant forms
\begin{align}
&P_{\mu}(IT)=\tfrac{1}{2}\int d^4p\delta((p_0)^2-E_p^2)\theta(p_0)\left[A^{\dagger}(IT;\vec{p})A(IT;\vec{p})+A(IT;\vec{p})A^{\dagger}(IT;\vec{p})\right]p_{\mu},\quad p_{\mu}=(p_0,p_1,p_2,p_3),
\label{9.4}
\end{align}
\begin{align}
&P_{\mu}(LF)=\int d^4p\delta(4p_+p_--F_p^2)\theta(p_+)\left[A^{\dagger}(LF;\vec{p})A(LF;\vec{p})+A(LF;\vec{p})A^{\dagger}(LF;\vec{p})\right]p_{\mu},\quad p_{\mu}=(p_+,p_1,p_2,p_-),
\label{9.5}
\end{align}
with the form for $P_{\mu}(IT)$ being given in \cite{Bjorken1965}.

We now apply $U(P_0,P_3)=\exp(ix^3P_0)\exp(ix^0P_3)$ to both sides of the $P_{\mu}(IT)$ equation given in (\ref{9.4}). Now the four momentum generators commute with each other according to $[P_{\mu},P_{\nu}]=0$. Thus $UP_{\mu}(IT)U^{-1}=P_{\mu}(IT)$, with the left-hand side being unchanged. On the right-hand side  we apply (\ref{8.24}) and obtain
\begin{eqnarray}
&&P_{\mu}(IT)=\tfrac{1}{2}\int d^4p\delta((p_0)^2-E_p^2)\theta(p_0)\left[A^{\dagger}(LF;\vec{p})A(LF;\vec{p})+A(LF;\vec{p})A^{\dagger}(LF;\vec{p})\right]p_{\mu}.
\label{9.6}
\end{eqnarray}
Changing the variables according to $p_0=p_++p_-$, $p_3=p_+-p_-$ yields
\begin{eqnarray}
&&P_{0}(IT)=\int d^4p\delta(4p_+p_--E_p^2)\theta(p_++p_-)\left[A^{\dagger}(LF;\vec{p})A(LF;\vec{p})+A(LF;\vec{p})A^{\dagger}(LF;\vec{p})\right](p_++p_-),
\nonumber\\
&&P_{3}(IT)=\int d^4p\delta(4p_+p_--E_p^2)\theta(p_++p_-)\left[A^{\dagger}(LF;\vec{p})A(LF;\vec{p})+A(LF;\vec{p})A^{\dagger}(LF;\vec{p})\right](p_+-p_-),
\nonumber\\
&&P_{i}(IT)=\int d^4p\delta(4p_+p_--E_p^2)\theta(p_++p_-)\left[A^{\dagger}(LF;\vec{p})A(LF;\vec{p})+A(LF;\vec{p})A^{\dagger}(LF;\vec{p})\right]p_i,\quad i=1,2.
\label{9.7}
\end{eqnarray}
Because of the delta function term we can replace the $\theta(p_++p_-)$ term by $\theta(p_+)$. We thus obtain
\begin{eqnarray}
&&P_{0}(IT)=\int d^4p\delta(4p_+p_--E_p^2)\theta(p_+)\left[A^{\dagger}(LF;\vec{p})A(LF;\vec{p})+A(LF;\vec{p})A^{\dagger}(LF;\vec{p})\right](p_++p_-),
\nonumber\\
&&P_{3}(IT)=\int d^4p\delta(4p_+p_--E_p^2)\theta(p_+)\left[A^{\dagger}(LF;\vec{p})A(LF;\vec{p})+A(LF;\vec{p})A^{\dagger}(LF;\vec{p})\right](p_+-p_-),
\nonumber\\
&&P_{i}(IT)=\int d^4p\delta(4p_+p_--E_p^2)\theta(p_+)\left[A^{\dagger}(LF;\vec{p})A(LF;\vec{p})+A(LF;\vec{p})A^{\dagger}(LF;\vec{p})\right]p_i,\quad i=1,2.
\label{9.8}
\end{eqnarray}
Finally, comparing with  (\ref{9.5}) we obtain
\begin{eqnarray}
P_0(IT)=P_+(LF)+P_-(LF),\quad P_3(IT)=P_+(LF)-P_-(LF),\quad P_1(IT)=P_1(LF),\quad  P_2(IT)=P_2(LF).
\label{9.9}
\end{eqnarray}
As we see, the relationship between the free theory instant-time and light-front momentum operators tracks the relationship between their eigenvalues. 

Now in constructing the all-order instant-time and light-front Lehmann representations in Sec. \ref{S8} the basic input was the existence of a complete set of all-order momentum eigenstates. Since that basis is complete the all-order momentum generators are given by 
\begin{eqnarray}
P_{\mu}(IT)=\sum|p^n(IT)\rangle p^n_{\mu}(IT)\langle p^n(IT)|, \quad P_{\mu}(LF)=\sum|p^n(LF)\rangle p^n_{\mu}(LF)\langle p^n(LF)|. 
\label{9.10}
\end{eqnarray}
With eigenvalues not changing under a unitary transformation, we obtain
\begin{align}
P_0(IT)=UP_0(IT)U^{-1}=U\sum| p^n(IT)\rangle p^n_{0}\langle p^n(IT)|U^{\dagger}=\sum|p^n(LF)\rangle (p^n_++p^n_-)\langle p^n(LF)|=P_+(LF)+P_-(LF).
 \label{9.11}
 \end{align}
The relations given in (\ref{9.9}) thus hold to all orders in interactions. In addition, we note that
from (\ref{9.9}) we obtain the  Lorentz invariant operator identity
\begin{eqnarray}
P^2_0(IT)- P^2_3(IT)-P^2_1(IT)-P^2_2(IT)=4P_+(LF)P_-(LF)-P^2_1(LF)-P^2_2(LF),
\label{9.12}
\end{eqnarray}
a relation that also holds to all orders in interactions. Now as written, each operator in (\ref{9.12}) is infinite-dimensional, with each possessing an infinite number of momentum eigenstates. However, when acting on any particular  set of momentum eigenstates $|p_{\mu}\rangle$ with eigenvalues $p_{\mu}$ that obey $p^2=m^2$ both sides of (\ref{9.12}) have eigenvalue $m^2$. 

Given (\ref{9.9}) and (\ref{9.11}), there initially appears to be a mismatch between the eigenstates of $P_0(IT)$ and $P_+(LF)$. However, for any timelike set of instant-time momentum eigenvalues we can Lorentz boost $p_1$, $p_2$ and $p_3$ to zero, to then leave $p_0=m$. If we impose this same $p_1=0$, $p_2=0$, $p_3=0$ condition on the light-front momentum eigenvalues we would set $p_+=p_-$, $p^2=4p_+^2=m^2$, and thus obtain $p_0=2p_+=m$. When written in terms of contravariant vectors with $p^{\mu}=g^{\mu\nu}p_{\nu}$ this condition takes the form $p^0=p^-$. Thus in the instant-time rest frame the eigenvalues of $P^0(IT)$ and $P^-(LF)$ coincide. In this sense then instant-time and light-front Hamiltonians are equivalent.

Finally, we note that assumption that underlies our analysis is that the theory is general coordinate invariant and that the momentum generators are Hermitian.  As long as we take renormalization counterterms to equally be general coordinate invariant and Hermitian, which we do, all of our results survive renormalization. Since in general in quantum theory invariance under translations entails  invariance under unitary transformations, we thus establish not only that instant-time quantization and light-front quantization are equivalent procedures, they are unitarily equivalent procedures.

\section{Final Comments}
\label{S10}

To conclude, we note that as well as establish the equivalence of instant-time quantization and light-front quantization for free quantum fields, through the use of the Lehmann representation we are able to show that our results immediately generalize to interacting theories. This has to be the case since perturbative interactions cannot change a Hilbert space, so once we show that the free instant-time and free light-front theories are in the same Hilbert space (as their (unequal time) commutators and anticommutators are related purely by kinematic coordinate transformations), it follows that the interacting theories are in the same Hilbert space too. In fact in general we note that because of the general coordinate invariance of quantum theory, any two directions of quantization that are related by a general coordinate transformation must describe the same theory. Since the transformation $x^0\rightarrow x^0+x^3=x^+$ is one such general coordinate transformation, it follows that to all perturbative orders light-front quantization is instant-time quantization. In fact for matrix elements of quantum field operators it was through the general coordinate invariance of both all-order Feynman diagrams and all-order path integrals (both c-number formulations of quantum field theory that only involve integrations over classical momenta or classical fields) that the all-order equivalence for matrix elements in both the instant-time and light-front cases was established \cite{Mannheim2019a,Mannheim2019b}. Finally, we note that while we have shown that light-front quantization is instant-time quantization, and while we have even shown that in the instant-time rest frame the instant-time and light-front Hamiltonians are equivalent, nonetheless the light-front formulation offers many computational benefits (see e.g. \cite{Brodsky:1997de}), and in that sense it could be preferred over instant-time quantization.

\begin{acknowledgments}
The author would like to thank Drs. S. J. Brodsky, P. Lowdon and L. Jin for helpful comments.
\end{acknowledgments}

\end{document}